\documentclass{article}
\usepackage{slashed,verbatim,subfigure}
\usepackage[numbers,sort&compress]{natbib}
\usepackage{amsmath}
\usepackage{amssymb}
\usepackage{appendix}
\usepackage{graphicx}
\usepackage{dcolumn}
\usepackage{bm}
\usepackage[colorlinks=true,linktocpage=true,
linkcolor=blue,citecolor=blue]{hyperref}
\usepackage[a4paper]{geometry}
\newcommand{\be}{\begin{eqnarray}}
\newcommand{\ee}{\end{eqnarray}}
\begin{document}
\title{Non perturbative and thermal dynamics of confined fields in dual QCD}
\author{H.C. Chandola$^{1}$,  Deependra Singh Rawat$^{1}$,
 H.C. Pandey$^{2}$, Dinesh Yadav$^{1}$ and H. Dehnen$^{3}$\\
$^{1}$Department of Physics (UGC-Centre of Advanced Study), Kumaun University, Nainital, India.\\
chandolahc@kunainital.ac.in, dsrawatphysics@kunainital.ac.in, dinoyadav02@rediffmail.com\\
$^{2}$Department of Physics, Birla Institute of Applied Sciences, Bhimtal, India\\ hempandey@birlainstitute.co.in\\
$^{3}$Fachbereich Physik, Universitat Konstanz, M677, 78457 Konstanz, Germany.\\
Heinz.Dehnen@uni-konstanz.de}
\maketitle

\abstract
{In order to study the detailed dynamics and associated non-perturbative features of QCD, a dual version of the color gauge theory based on the topologically viable ho-
mogeneous fiber bundle approach has been analysed taking into account its magnetic symmetry structure. In the dynamically broken phase of magnetic symmetry, the associated
flux tube structure on a $S^{2}$-sphere in the magnetically condensed state of the dual QCD vacuum has been analyzed for the profiles of the color electric field using flux quantization and stability conditions. The color electric field has its intimate association with the vector mode of the magnetically condensed QCD vacuum and such field configurations have been analyzed to show that the color electric flux gets localized towards the poles for a large sphere case while it gets uniformly distributed for the small sphere case in the infrared sector of QCD. The critical flux tube densities
have been computed for various couplings and are shown to be in agreement with that for lead-ion central collisions in the near infrared sector of QCD. The possible annihila-
tion/unification of flux tubes under some typical flux tube density and temperature conditions in the magnetic symmetry broken phase of QCD has also been analyzed and shown
to play an important role in the process of QGP formation. The thermal variation of the profiles of the color electic field is further investigated which indicates the survival of flux tubes even in the thermal domain that leads the possibility of the formation of some exotic states like QGP in the intermedate regime during the quark-hadron phase transition.}

\section{Introduction}

\hspace{0.5cm}It is widely believed that the Quantum chromodynamics as a non-Abelian theory of gauge fields can be very well used for the fundamental description of strong interactions \cite{Gross}-\cite{Polit} between quarks and gluons in its high energy sector. Apart from this, it is also known to exhibit many important non-perturbative features, like confinement, chiral symmetry breaking, mass spectrum of the physical states etc as far as its low energy sector \cite{Stack}-\cite{Kronfield} is concerned. The color confinement in QCD is naturally one of the most challenging issue and it is directly linked with the physical spectrum of the theory. In low energy sector of QCD, there have been many conjectures proposed on the mechanism of color confinement. In this context, Nambu \cite{N74} and others \cite{H78}-\cite{Bak91} have argued that the monopole condensation leads dual Meissner effect that could provide a way for the color confinement in QCD in a manner similar to the ordinary superconductor where magnetic flux confinement occurs due to the Meissner effects. To ensure the color confinement through dual Meissner effect \cite{Suganuma},\cite{Suz88},\citep{Sug95}, however, one needs color monopoles as the most essential degrees of freedom. The appearence of color monopoles in QCD was proposed by 't Hooft \cite{H81} based on the Abelian gauge fixing that reduces SU(N) gauge theory to $U(1)^{N_{c}-1}$ with color monopoles as a topological excitations of QCD vacuum. However, the mechanism of monopole condensation in QCD vacuum remains far from clear. Afterwards, the chromo-electric flux between static quarks is then squeezed into string or tube like vortex in the superconductivity and such flux tube representation of hadrons leads to a lineraly rising potential between static quarks at zero temperature, characterizing the color confinement, and is strongly supported by the recent lattice QCD simulation techniques \cite{K6}-\cite{C14}. Thus, the apparent dominance of color monopole condensation seems to provide a viable explanation of color confinement and related features of the infrared sector of QCD. However, inspite of the great success of lattice calculation, it is tempting to develop an approach based on first principle of QCD, which may provide us with a clearer understanding of the physical picture of dual QCD vacuum, associated mechanism and allows us to perform some analytical calculation also. Therefore, an analytical effective theory of nonperturbative QCD is highly desirable which may incorporate the topological aspects, in addition to various glueballs and their interactions, which are the most practical degrees of freedom for the study of the QCD phase transition. In this direction, a topologically effective magnetic symmetry based dual gauge formulation \cite{hcc1}-\cite{hcc7} might prove a parallel to explain the mysterious confining behaviour and associated features of quarks inside the hadronic jails. Furthermore, the collective behavior of color isocharges at finite temperature and density has also attracted a great interest in the theoretical as well as experimental investigations of QCD phase diagram \cite{bdm}. In this way, the dual QCD formulation is expected to provide the relevant understanding of various non-perturbative aspects of QCD, the QCD phase transition and the possible formation of an intermediatery state like  quark-gluon plasma (QGP) \cite{shk1}-\cite{abbas} or any other exotic states under the extreme conditions of density and temperature\cite{satz83}-\cite{lizardo02} which might also be relevant to uncover the many hidden aspects of the primordial structure of universe. In the present study, the mechanism of color confinement, associated confined field profiles and the QCD phase transition are studied by investigating the flux tube structure associated with dual QCD formulation based on topologically viable magnetic symmetry approach and analysing on energetic grounds. The typical  dual QCD parameters of quark-hadron phase transition have been evaluated and analysed numerically. The study is further extended to investigate the profiles of color electric field at different hadronic scales using energy minimization and stabilty conditions. The critical flux tube densities have been computed and analyzed for QGP formation in heavy-ion collision events. The field profiles have further been investigated for better understanding of the QCD phase transition under varying thermal conditions.\\
\section{Magnetic Symmetry structure and confining features of Dual QCD}
In order to analyze the typical non-perturbative features of the color gauge theory, let us briefly review the dual formulation of QCD \cite{bak1}-\cite{hcc7} in terms of the magnetic symmetry and the associated topological structure of dual QCD vacuum. It is well known that the non-Abelian theory of gauge fields can be viewed \cite{cho1}-\cite{hcc7} as the Einstein theory of gravitation in a higher-dimensional unified space that allows the introduction of some additional internal isometries. In this connection, the magnetic symmetry may be introduced as a set of self-consistent Killing vector fields of the internal space which, while keeping the full gauge degrees of freedom intact, restricts and reduces some of the dynamical degrees of freedom of the theory. For the case of quark color symmetry, it, in turn, may be shown to establish a dual dynamics between the color isocharges and the topological charges of the underlying gauge group. For the simplest choice of the gauge group $G \equiv SU(2)$ with its little group $H \equiv U(1)$, the gauge covariant magnetic symmetry condition may be expressed \cite{hcc1}-\cite{hcc7} in the following form,\\
\begin{equation}
D_{\mu}\hat{m} = 0~~~\Rightarrow~~~(\partial_{\mu}+g{\bf{W_{\mu}}}\times)\hat{m}=0,
\end{equation}
where $\hat{m}$ is a scalar multiplet that belongs to the adjoint representation of the gauge group G and ${\bf{W_{\mu}}}$ is the associated gauge potential of the underlying gauge group G. The condition (1) thus implies that the magnetic symmetry imposes a strong constraint on the metric as well as connection and may, therefore, be regarded as the symmetry of the potentials. The monopoles, therefore, emerge as the topological objects associated with the elements of the second homotopic group $\prod_{2}(G/H)$. The typical potential satisfying the condition (1) is identified as,\\
\begin{equation}
{\bf{W_{\mu}}}=A_{\mu}\hat{m}-g^{-1}(\hat{m}\times\partial_{\mu}\hat{m})
\end{equation}
where $\hat{m}.{\bf{W_{\mu}}}= A_{\mu}$ is the color electric potential unrestricted by magnetic symmetry, while the second term is completely determined by magnetic symmetry and is topological in origin. Thus, the virtue of the magnetic symmetry is that it can be used to describe the topological structure of gauge symmetry and the multiplet $\hat{m}$ may then be viewed as defining the homotopy of the mapping $\prod_{2}(S^{2})$ on $\hat{m}:S_{R}^{2}\rightarrow S^{2}=SU(2)/U(1)$. The imposition of magnetic symmetry on the gauge group thus brings the topological structure into the dynamics explicitly. The associated field strength corresponding to the potential (2) is then given by,\\
\begin{equation}
{\bf{G_{\mu\nu}=W_{\nu,\mu}-W_{\mu,\nu}+gW_{\mu}\times W_{\nu}}}\equiv(F_{\mu\nu}+B_{\mu\nu}^{(d)})\hat{m}
\end{equation}
where,~~$F_{\mu\nu}=A_{\nu,\mu}-A_{\mu,\nu}$ and $B_{\mu\nu}^{(d)} = -g^{-1}\hat{m}.(\partial_{\mu}\hat{m}\times\partial_{\nu}\hat{m}) = B_{\nu,\mu}-B_{\mu,\nu}$. The second part ($B_{\mu}$), fixed completely by $\hat{m}$, is thus identified as the magnetic potential associated with the topological monopoles and the fields thus appear in a completely dual symmetric way.\\
In order to explain the dynamics of the resulting dual QCD vacuum and its implications on confinement mechanism, let us start with the SU(2) chromodynamic Lagrangian with a quark doublet source $\psi(x)$, as given by,\\
\begin{equation}
{\bf{\mathcal{L}}}=-\frac{1}{4}{\bf{G_{\mu\nu}^{2}}}+\bar{\psi}(x)i\gamma^{\mu}{\bf{D_{\mu}}}\psi(x)-m_{0}\bar{\psi}(x)\psi(x).
\end{equation}
In addition, in order to avoid the problems due to the point like structure and the singular behavior of the potential associated with monopoles we use the dual magnetic potential $B_{\mu}^{(d)}$ coupled to a complex scalar field $\phi(x)$. 
Taking these considerations into account, the modified form of the dual QCD Lagrangian (4) in quenched approximation may be expressed as,\\
\begin{equation}
\mathcal{L}_{m}^{(d)}=-\frac{1}{4}B_{\mu\nu}^{2}+\vert[\partial_{\mu}+i\frac{4\pi}{g}B_{\mu}^{(d)}]\phi \vert^{2}-V(\phi^{\ast}\phi)
\end{equation}
with $V(\phi^{\ast}\phi)$ as a proper effective potential which is fixed by the requirements of the ultraviolet finiteness and infrared instability of the Lagrangian as given by Coleman and Weinberg \cite{CW} by using the single loop expansion technique as,\\
\begin{equation}
V(\phi^{\ast}\phi)=\frac{24\pi^{2}}{g^{4}}[\phi_{0}^{4}+(\phi^{\ast}\phi)^{2}\lbrace2ln\frac{\phi^{\ast}\phi}{\phi_{0}^{2}}-1\rbrace].
\end{equation}
The effective potential is reliable in the far infrared region where coupling becomes very intense ($\alpha_{s}\rightarrow 1$) and confinement is strongly enforced. Since, in the present case, we are also interested to investigate phase transition in dual QCD vacuum, the use of an effective potential reliable in relatively weak coupling near-infrared regime is naturally desired and therefore, the appropriate choice for inducing the dynamical breaking of magnetic symmetry is the quartic potential of the following form,\\ 
\begin{equation}
V_{pt}(\phi^{\ast}\phi) = 3\lambda\alpha_{s}^{-2}(\phi^{\ast}\phi-\phi_{0}^{2})^{2}.
\end{equation}
In order to analyze the nature of magnetically condensed vacuum and associated flux tube structure, let us investigate the field equations led by the Lagrangian (5) in the following form,\\
\begin{equation}
(\partial^{\mu}-i\frac{4\pi}{g}B^{(d)\mu})(\partial_{\mu}+i\frac{4\pi}{g}B_{\mu}^{(d)})\phi+6\lambda\alpha_{s}^{-2}(\phi^{\ast}\phi-\phi_{0}^{2})\phi=0,
\end{equation}

\begin{equation}
\partial^{\nu}B_{\mu\nu}+i\frac{4\pi}{g}(\phi^{\ast}\buildrel \leftrightarrow \over\partial_{\mu}\phi)-8\pi\alpha_{s}^{-1}B_{\mu}^{(d)}\phi\phi^{\ast}=0.
\end{equation}

The unusual features of dual QCD vacuum responsible for its non-perturbative behaviour may become more transparent if we start with the Neilsen and Olesen \cite{NO} interpretation of vortex like solutions. It leads to the possibility of the existence of the monopole pairs inside the superconducting vacuum in the form of thin flux tubes that may be responsible for the confinement of any colored fluxes. Under cylindrical symmetry $(\rho,\varphi,z)$ and the field ansatz given by,\\

\begin{center}
$B_{\varphi}^{(d)}(x)=B(\rho),~~~B_{0}^{(d)}=B_{\rho}^{(d)}=B_{z}^{(d)}=0$
\end{center}

\begin{equation}
and~~\phi(x)=exp(in\varphi)\chi(\rho)~~(n=0,\pm1,\pm2,--),
\end{equation}
the field equations (8) and (9) are transformed to the following form,\\
\begin{equation}
\frac{1}{\rho}\frac{d}{d\rho}\biggl(\rho\frac{d\chi}{d\rho}\biggr)-\biggl[\biggl(\frac{n}{\rho}+(4\pi\alpha_{s}^{-1})^\frac{1}{2}B(\rho)\biggr)^{2}+6\lambda\alpha_{s}^{-2}(\chi^{2}-\phi_{0}^{2})\biggr]\chi(\rho) = 0,
\end{equation}
\begin{equation}
\frac{d}{d\rho}\biggl[\rho^{-1}\frac{d}{d\rho}(\rho B(\rho))\biggr]-(16\pi\alpha_{s}^{-1})^{\frac{1}{2}}\biggl(\frac{n}{\rho}+\frac{4\pi}{g}B(\rho)\biggr)\chi^{2}(\rho)=0.
\end{equation}
Further, with these considerations, the form of the color electric field in the z-direction is given by,\\
\begin{equation}
E_{m}(\rho)=-\frac{1}{\rho}\frac{d}{d\rho}(\rho B(\rho))
\end{equation}
For a more convenient representation, we use the following dimensionless parameter,
\begin{equation}
r=2\sqrt{3\lambda}\alpha_{s}^{-1}\phi_{0}\rho,~~F(r)=(4\pi\alpha_{s}^{-1})^{\frac{1}{2}}\rho B(\rho),~~H(r)=\phi_{0}^{-1}\chi(\rho)
\end{equation}
so that the field equations (11) and (12) (with $\lambda=1$) are reduced to a more simpler form as,
\begin{equation}
H^{''}+\frac{1}{r}H^{'}+\frac{1}{r^{2}}(n+F)^{2}+\frac{1}{2}H(H^{2}-1)=0,
\end{equation}
\begin{equation}
F^{''}-\frac{1}{r}F^{'}-\alpha(n+F)H^{2}=0,
\end{equation}
where, $\alpha=2\pi\alpha_{s}/3\lambda$ and the prime stands for the derivative with respect to r. Using the asymptotic boundary conditions given by, $F\rightarrow~-n$, $H\rightarrow~1$ as $r\rightarrow\infty$, alongwith equation (14), the asymptotic solution for the function $F$ may be obtained in the following form,\\
\begin{equation}
F(\rho) = -n + C\rho^{\frac{1}{2}}exp(-m_{B}\rho)
\end{equation}
where, $C=2\pi B(3\sqrt{2}\lambda g^{-3}\phi_{0})^{\frac{1}{2}}$ and $m_{B}=(8\pi\alpha_{s}^{-1})^{\frac{1}{2}}\phi_{0}$ is the mass of the magnetic glueballs which appears as vector mode of the magnetically condensed QCD vacuum. Since, the function $F(\rho)$ is associated with the color electric field given by equation (13) through gauge potential $B(\rho)$, it indicates the emergence of dual Meissner effect leading to the confinement of the color isocharges in the magnetically condensed dual QCD vacuum.

\section{Critical parameters of phase transition in dual QCD vacuum}
The formation of color electric-flux tubes and the confinement of color charges can be visualized more effectively on the energetic grounds by evaluating the energy per unit length of the flux tube configuration governed by the field equations (15) and (16) and the Lagrangian (5), which is obtained in the following form,\\
\begin{center}
$k = 2\pi\phi_{0}^{2}\int_{0}^{\infty}rdr\biggl[\frac{6\lambda}{g^{2}}\frac{(F')^{2}}{r^{2}}+\frac{(n+F)^{2}}{r^{2}}H^{2}+(H')^{2}+\frac{(H^{2}-1)^{2}}{4}\biggr]$,
\end{center}
which in view of equation (14) reduces to,\\
\begin{equation}
k = 2\pi\int_{0}^{\infty}\rho d\rho\biggl[\frac{n^{2}g^{2}}{32\pi^{2}}\biggl(\frac{1}{\rho}\frac{dF}{d\rho}\biggr)^{2}+\frac{n^{2}}{\rho^{2}}F^{2}(\rho)\chi^{2}(\rho)
+\biggl(\frac{\partial\chi}{\partial\rho}\biggr)^{2}+\frac{48\pi^{2}}{g^{4}}\biggl(\chi^{2}(\rho)-\phi_{0}^{2}\biggr)^{2}\biggr]
\end{equation}
It, in turn, plays an important role in the phase structure of QCD vacuum, if we take the multi-flux tube system on a $S^{2}$-sphere with periodically distributed flux tubes and intoduce a new variables $R$ on $S^{2}$ and express it as, $\rho=Rsin\theta$. As a result, a number of flux tubes considered here inside a hadronic sphere of radius $R$ pass through the two poles of the hadronic sphere. Under such prescription , the flux tube solution governed by equation (11) and (12) corresponds to the case of large R limit $(R\rightarrow\infty)$ such that $R$$>>$$\rho$ and $\theta\rightarrow 0$. With these considerations, the finite energy expression given by above equation (18) may be reexpressed as,\\
\begin{equation}
k=\varepsilon_{C}+\varepsilon_{D}+\varepsilon_{0}~~with~~\varepsilon_{C}=k_{C}R^{2},~~\varepsilon_{D}=k_{D}R^{-2},~~\varepsilon_{0}=k_{0}
\end{equation}
where the functions $k_{C}$,~$k_{D}$ and $k_{0}$ are given by,
\begin{equation}
k_{C}=\frac{6\pi}{\alpha_{s}^{2}}\int_{0}^{\pi}[\chi^{2}(\theta)-\phi_{0}^{2}]^{2}sin\theta~d\theta,
\end{equation}
\begin{equation}
k_{D}=\frac{n^{2}\alpha_{s}}{4}\int_{0}^{\pi}\frac{1}{sin\theta}\biggl(\frac{\partial F}{\partial\theta}\biggr)^{2}d\theta,
\end{equation}
\begin{equation}
k_{0}=2\pi\int_{0}^{\pi}\biggl[\frac{n^{2}F^{2}(\theta)\chi^{2}(\theta)}{sin\theta}+sin\theta\biggl(\frac{\partial\chi}{\partial\theta}\biggr)^{2}\biggr]d\theta.
\end{equation}
The energy expression (19) provides an straightforward description of the behavior of QCD vacuum at different energy scales. At large hadronic distances, the first term ($\varepsilon_{C}$) in equation (19) dominates which increases at increasing hadronic distances and gets minimized when the monopole field picks up its non-zero vacuum expectation value which incidently acts as an order parameter to indicate the onset of the dynamical breaking of magnetic symmetry. The associated magnetic condensation of QCD vacuum then forces the color electric field to localize in the form of the thin flux tubes extending from $\theta$ = 0 to $\theta$ = $\pi$ and the QCD vacuum is ultimately pushed to the confining phase. Further, the energy expression (18) has its own implications for the evaluation of critical parameters of phase transition and their numerical computation then leads to a deep significance for the validity of field decomposition formulation of dual QCD. For the computation of such critical factors, we proceed by evaluating the functions associated with the expression (19) in terms of basic free parameters of the theory (viz., $\alpha_{s}$~and~$m_{B}$) in the following way,\\ 
\begin{equation}
k_{C}=\frac{6\pi}{\alpha_{s}^{2}}\int_{0}^{\pi}[\chi^{2}(\theta)-\phi_{0}^{2}]^{2}sin\theta~d\theta~~~~\Rightarrow~~~~k_{C}=\frac{3m_{B}^{4}}{16\pi},
\end{equation}
which shows the vector mode mass of the magnetically condensed vacuum plays a crucial role in the confining phase of the QCD vacuum. On the other hand, the component of energy  dominant over relatively short hadronic distances ($\varepsilon_{D}$), expressed in terms of the function $k_{D}$ given by equation (21), may also be evaluated in terms of free parameters of the theory in the form as given below:
\begin{equation}
k_{D} = \pi~R^{4}\int_{0}^{\pi}E_{m}^{2}(\theta)sin\theta~d\theta,
\end{equation} 
where,
\begin{equation}
E_{m}(\theta)=\frac{ng}{4\pi~R^{2}sin\theta}\frac{\partial F}{\partial\theta}.
\end{equation}
Using equations (24) and (25) alongwith the flux quantization condition given by\\
\begin{equation}
\int\rho~E_{m}(\rho)d\rho=\frac{ng}{4\pi}
\end{equation}
then leads to,
\begin{equation}
k_{D}=\frac{n^{2}g^{2}}{8\pi}=\frac{1}{2}n^{2}\alpha_{s}.
\end{equation}
The contributions, in terms of the energy functions $k_{C}$ and $k_{D}$ may be used to compute the critical parameters of phase transition from magnetically dominated phase to electrically dominated one. For this purpose, let us evaluate the ratio of $\varepsilon_{D}$ and $\varepsilon_{C}$ which is obtained as,
\begin{equation}
\frac{\varepsilon_{D}}{\varepsilon_{C}}=\eta\biggl(\frac{1}{R^{4}}\biggr),~~~~where,~~~~\eta=\frac{k_{D}}{k_{C}}=\frac{8\pi~n^{2}\alpha_{s}}{3m_{B}^{4}}.
\end{equation}
For the confinement-deconfinement phase transition, we have $\frac{\varepsilon_{D}}{\varepsilon_{C}}$=1 and $R=R_{c}$ which leads to the critical radius of phase transition in the following way, 
\begin{equation}
R_{c}=\left(\frac{8}{3}\pi n^{2}\alpha_{s}\right)^{\frac{1}{4}}m_{B}^{-1}.
\end{equation}
The corresponding critical density of the flux tubes ($d_{c}$) inside the hadronic sphere is given by,
\begin{equation}
d_{c}=\frac{1}{2\pi R_{c}^{2}}=\left(\frac{32}{3}\pi^{3}n^{2}\alpha_{s}\right)^{-\frac{1}{2}}m_{B}^{2}.
\end{equation}
These parameters are extreamly important in exploring the underlying mechanism and nature of QCD phase transition. The equations (29) and (30) exhibit that the critical radius and critical density of phase transition are clearly expressible in terms of free parameters of the QCD vacuum. In view of the running nature of QCD coupling constant, we can estimate these critical factors associated with the QCD vacuum in its infrared sector using the numerical estimations of glueball masses \cite{cho1}. For instance, for the optimal value of ($\alpha_{s}$) as $\alpha_{s}\equiv 0.12$ with the glueball masses $m_{B}=2.102 GeV$ and $m_{\phi}=4.205 GeV$, equations (29) and (30), lead to,\\
\begin{center}
$R_{c}=0.094~fm~~~and~~~d_{c}=18.003~fm^{-2}.$
\end{center}
The deconfinement phase transition in the magnetically condensed QCD vacuum therefore appears around the above-mentioned critical values for a typical coupling of $\alpha_{s}=0.12$ in the near infrared sector of QCD. In this case, for $R_{c} \equiv$ 0.094 fm, the flux tube density acquire its critical value of 18.003 $fm^{-2}$ and the first part of the energy expression (19) dominates which demonstrates the confinement of color particles in the magnetically rich QCD vaccum. However, below $R_{c}=0.094$ fm, the quarks and gluons appear as free states and the system stands near the boundary of the perturbative phase where the second part of the expression (19) becomes dominant leading to the deconfinement of color isocharges. The flux tube density in this sector increases sharply and with sufficiently dense flux tube system, the flux tube annihilation may takes place which then leads to the generation of dynamical quarks and gluons. The gluon self-interactions are then expected to play a major role in the thermalization of QCD system and create an intermediatery state of quark-gluon plasma (QGP). As a result of such flux tube melting in the high momentum transfer sector of QCD vacuum, the system is expected to evolve with an intermediatory QGP phase. In addition, in the deep infrared sector of QCD with higher couplings, ($\alpha_{s}$) e.g. 0.24, 0.48 and 0.96, a considerable enhancement in the critical radius (0.152 fm, 0.226 fm and 0.356 fm) is observed which leads to a formidable depletion in critical density (7.22,3.10,1.24 $fm^{-2}$~respectively) of flux tubes in QCD vacuum. As shown by color electric field profiles, the color electric field given by equation (25) obtained for the asymptotic solution (large $R$) leads to uniform field when $R$ decreases to approach its critical value $R_{c}$ where the flux tube density acquires its critical value $d_{c}$ around which a number of flux tubes are expected to get melted . These critical flux tube densities may be compared to those produced in case of central heavy-ion collisions for the creation of QGP. The flux tube densities for such collisions \cite{sug96} for the case of one flux tube per nucleon-nucleon hard collision, is given by,\\
\begin{equation}
d_{c} = \frac{9A^{2/3}}{4\pi R_{0}^{2}}
\end{equation} \\
where A is the mass number of heavy-ions and $R_{0}$ is nucleus radius parameter (1.2 fm). For the case of Pb-Pb central collisions (A=208), it leads to a value of 17.5 $fm^{-2}$, which is in very good agreement with that given by equation (30) of the present dual QCD model and indicates the vital role played by the flux tube number density and associated monopole condensate in QGP phase transition in QCD.\\
\section{Confined Field Configurations in dual QCD vacuum}

\begin{figure}
\resizebox{0.40\textwidth}{!}{
\includegraphics{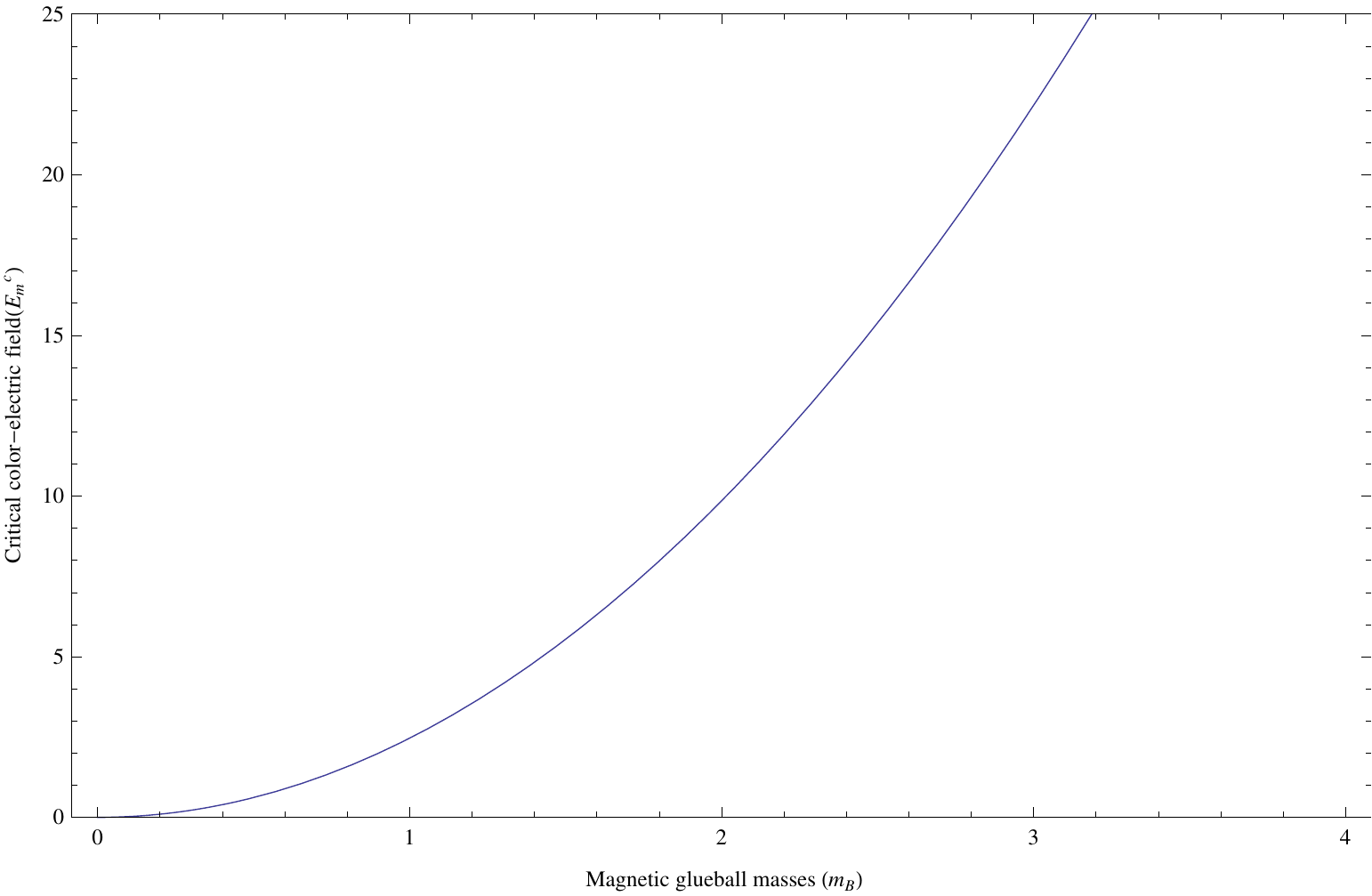}}
\caption{(Color online) Variartion of critical color electric field $E_{m}(critical)$ with $m_{B}$.}
\end{figure}

\begin{figure}[htp]
\centering
\begin{tabular}{cc}
\includegraphics[width=80mm]{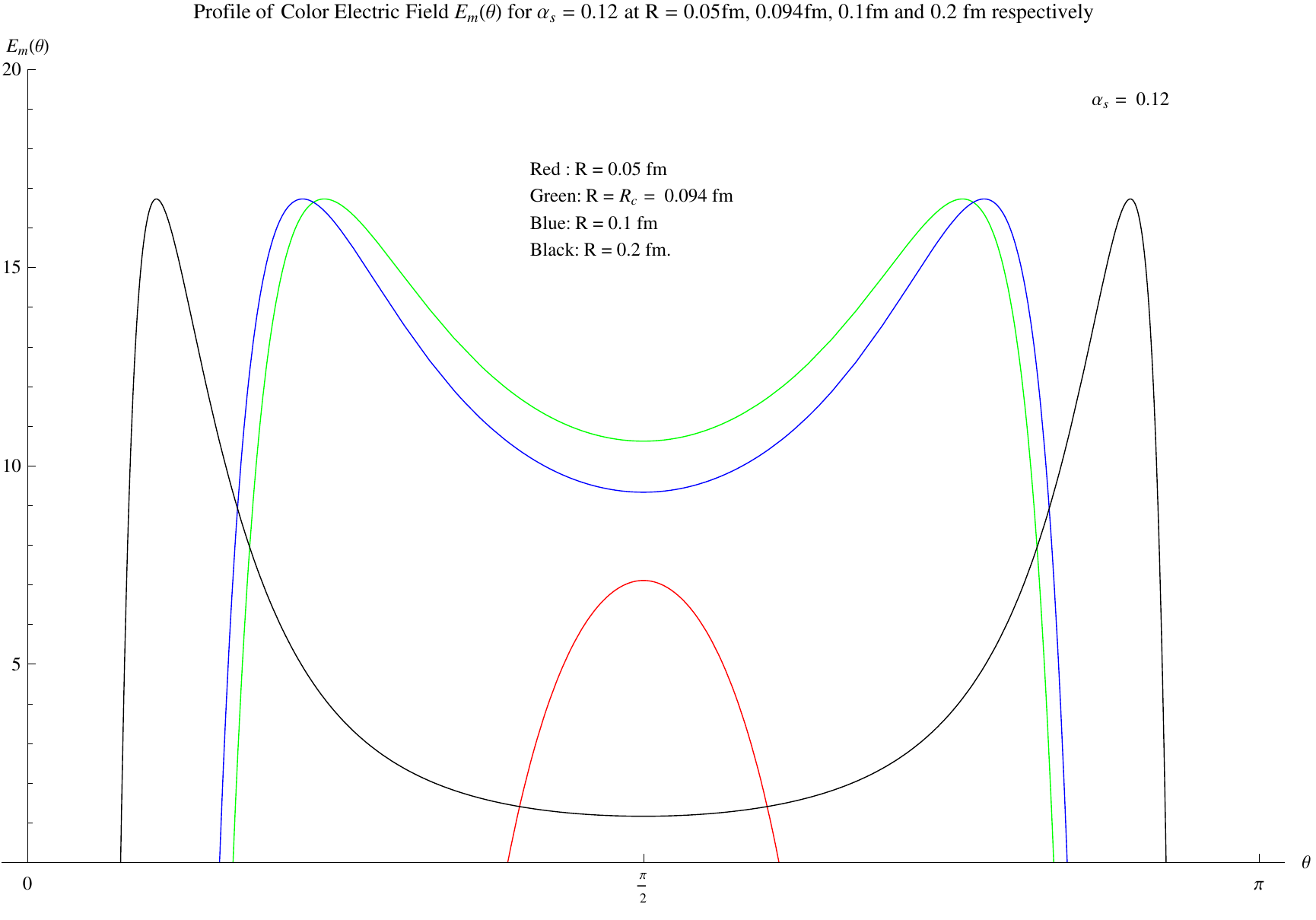}\\
\includegraphics[width=80mm]{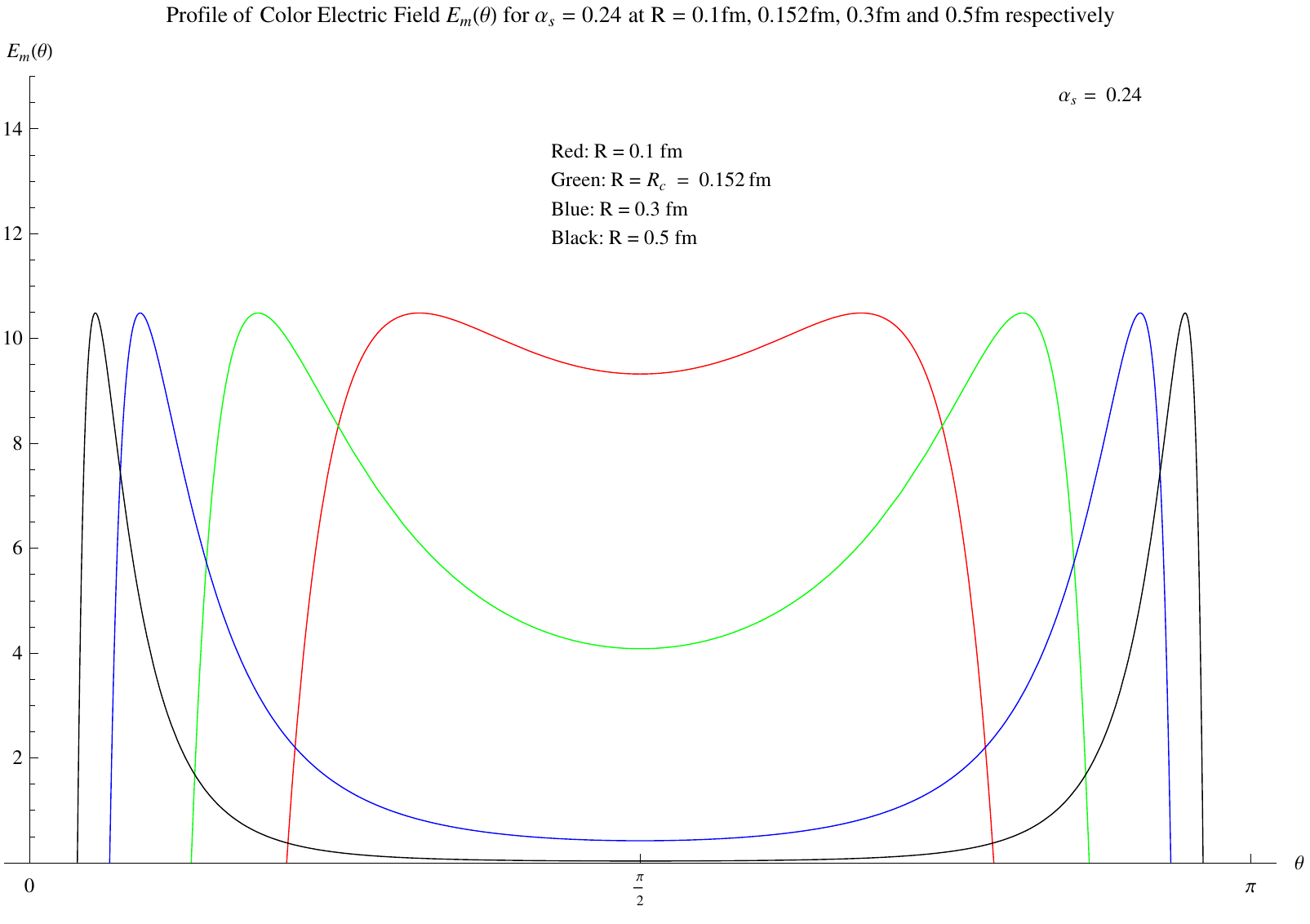}\\
\includegraphics[width=80mm]{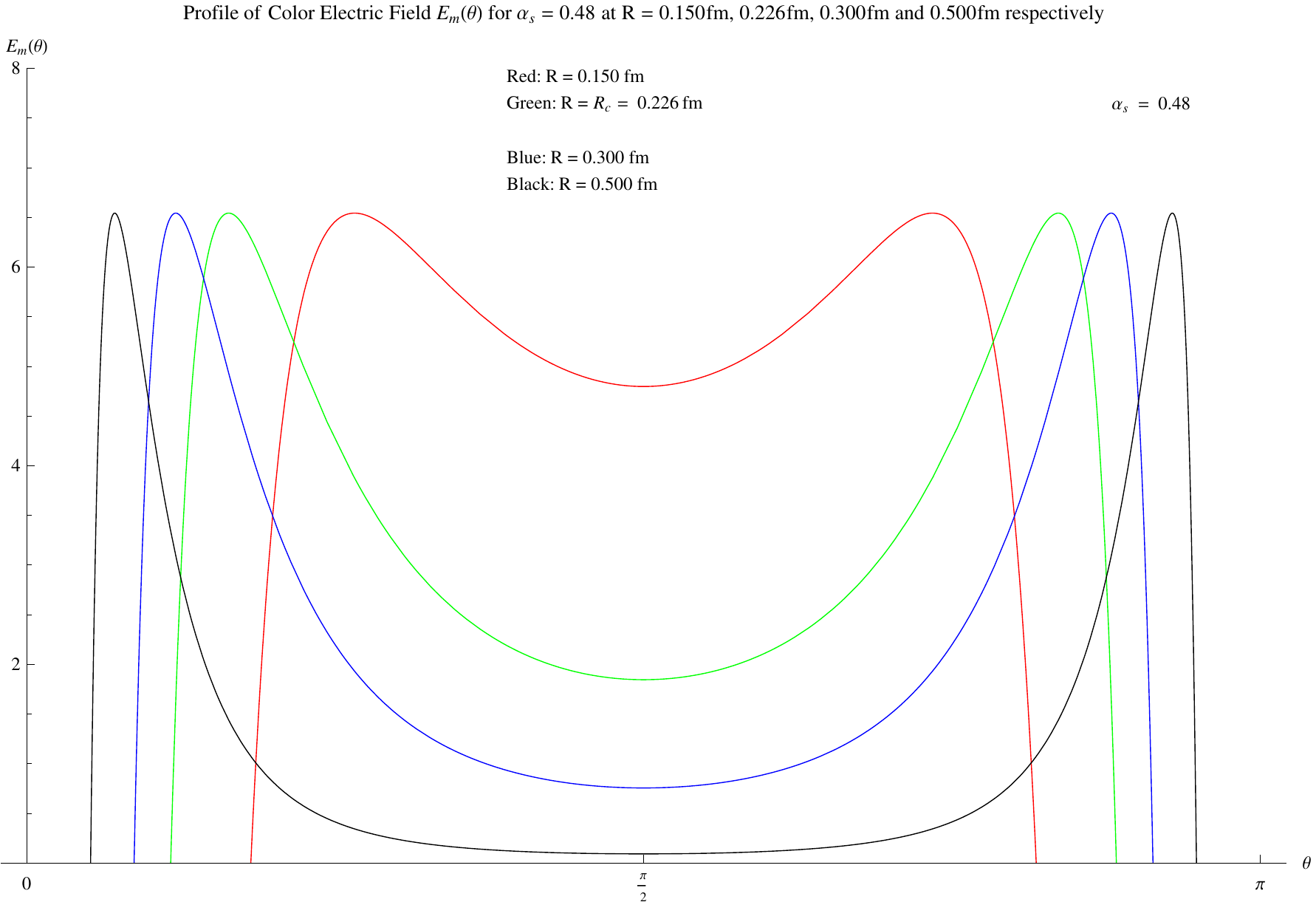}\\
\includegraphics[width=80mm]{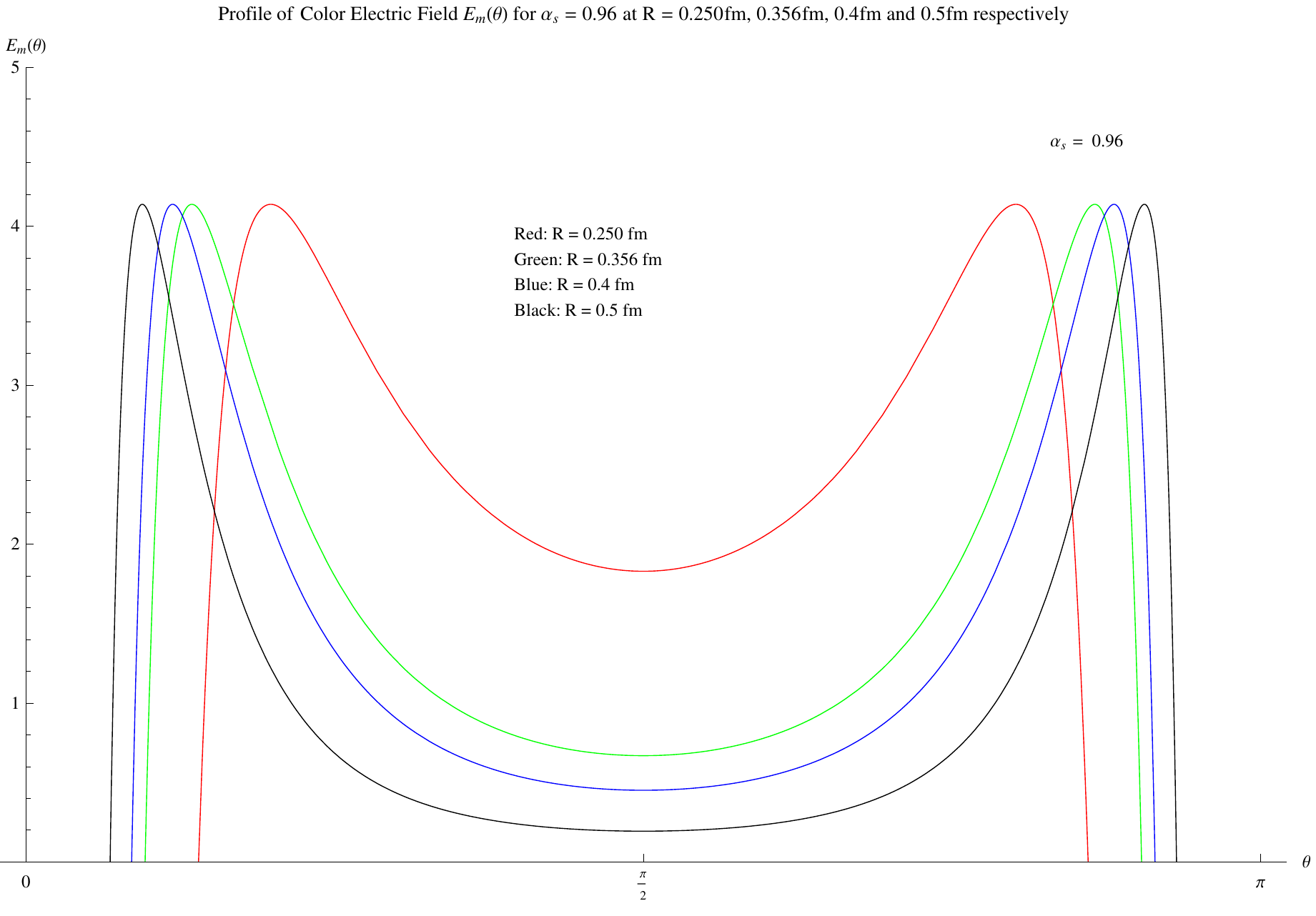}\\
\end{tabular}
\caption{(Color online) Profiles of color electric field $E_{m}(\theta)$ as a function of
$\theta$ for different $R$ at $\alpha_{s}=0.12, 0.24, 0.48~~and~~0.96~~respectively.$}
\end{figure}

In order to further discuss the phase structure of QCD vacuum in terms of the critical parameters, let us extend our study to the profiles of the color electric field in the full infrared sector of dual QCD. Using the prescription for the color electric field and potential as given in the previous section, the total electric flux penetrating the area ($S$) surrounded by a closed loop around the upper sphere of radius $R$ is given by,\\
\begin{equation}
\phi=\int_{S}{\bf{{E}_{m}}}.{\bf{dS}} \equiv\oint {\bf{B^{(d)}.dl}} = -2\pi\int_{0}^{\infty}\rho E_{m}(\rho)d\rho =-2\pi R^{2}\int_{0}^{\frac{\pi}{2}}E_{m}(\theta)sin\theta d\theta
\end{equation}
Using the flux quantization condition along with the substitution of the variable, $cos\theta=p$, it yields\\
\begin{equation}
\int_{0}^{\frac{\pi}{2}}E_{m}(\theta) sin\theta d\theta =\int_{0}^{1}E_{m}(p)dp = \frac{ng}{4\pi R^{2}} \equiv N.
\end{equation}
This alongwith equation (19) may be used to evaluate the flux tube energy component $\varepsilon_{D}$ in the following form\\
\begin{equation}
\varepsilon_{D}=\pi R^{2}\left[\int_{0}^{1}\lbrace E_{m}(p)-N\rbrace^{2}dp+N^{2}\right].
\end{equation}
The energy minimization condition then leads to,\\
\begin{equation}
E_{m}(p)=N=\frac{ng}{4\pi R^{2}},
\end{equation}
so that the color electric field is distributed uniformly in the deconfinement region where $\varepsilon_{D}$ dominates. The critical value of such color electric field at the boundary of phase transition from the deconfined phase to the confined phase may then be obtained by using $R=R_{c}$, in the following form,\\
\begin{equation}
E_{m}^{c}= \frac{ng}{4\pi R_{c}^{2}}=\frac{1}{4\pi}\sqrt{\frac{3}{2}}m_{B}^{2},
\end{equation}
which leads to its numerical value as $E_{m}^{c}(fm^{-2})$ = 23.16, 11.03, 5.81, 3.80 and 2.16 at strong couplings $\alpha_{s}$ = 0.048, 0.12, 0.24, 0.48 and 0.96 rspectively in the full infrared sector of QCD. The variation of such critical color electric field at the phase transition boundary, as depicted in figure (1), shows a large reduction in the color electric flux spread out in deep infrared sector on one hand and a considerable enhancement in its value on the other, in the transitional region where a number of flux tubes are expected to lead a homogeneous QGP as a result of their annihilation.\\

Furthermore, the general form of color-electric field may be evaluated by using equation (13) alongwith the asymptotic solution (17) and is given as,\\
\begin{equation}
E_{m}(\rho)=\frac{ngC}{8\pi\rho^\frac{3}{2}}(1-2m_{B}\rho)exp(-m_{B}\rho)
\end{equation}
For the case of multi-flux tube system on the $S^{2}$-sphere, the flux-tube are periodically distributed over the sphere of radius $R$ and the associated color electric field passing vertically on the surface of sphere is obtained as,\\
\begin{equation}
E_{m}(\theta)= \tilde{E}_{m} exp(-Rm_{B}sin\theta),
\end{equation}
\begin{center}
where,~~$\tilde{E}_{m}=\frac{nC\alpha_{s}^{1/2}}{4\pi^{1/2}R^{3/2}sin^{3/2}\theta}(1-2Rm_{B} sin\theta).$
\end{center}
The profile of such color electric field as a function of the polar angle $\theta$ for different values of radius ($R$) at different $\alpha_{s}$ in the infrared sector of QCD has been presented by a 2-d graphics given by figure (2). It clearly shows that, in the infrared sector of QCD, for a large sphere enclosing the flux tubes, the color electric flux gets localized or spreaded around the poles ($\theta$ = 0~and~$\pi$) while its gets uniformly distributed for the small sphere case and acquires a constant value at the critical radius $R_{c}$ as given by equation (29). Similar results can be drawn from the 3-d graphics of fig. 3 for the color electric field as a function of the polar angle $\theta$ and radius $R$ for different values of coupling constant $\alpha_{s}$. In these graphics with the increase in radius of sphere in the infrared sector of QCD, it demonstrate, the reduction and drifting of maxima of color electric flux towards the higher sphere radii and consequently a reduction in flux tube density in the far infrared sector of QCD. In the near-infrared sector, however, the increase in flux tube density may lead to the annihilation of the neighbouring flux tubes and a large homogeneous QGP formation in the central region is expected.

\begin{figure}[htp]
\centering
\begin{tabular}{cc}
\includegraphics[width=80mm]{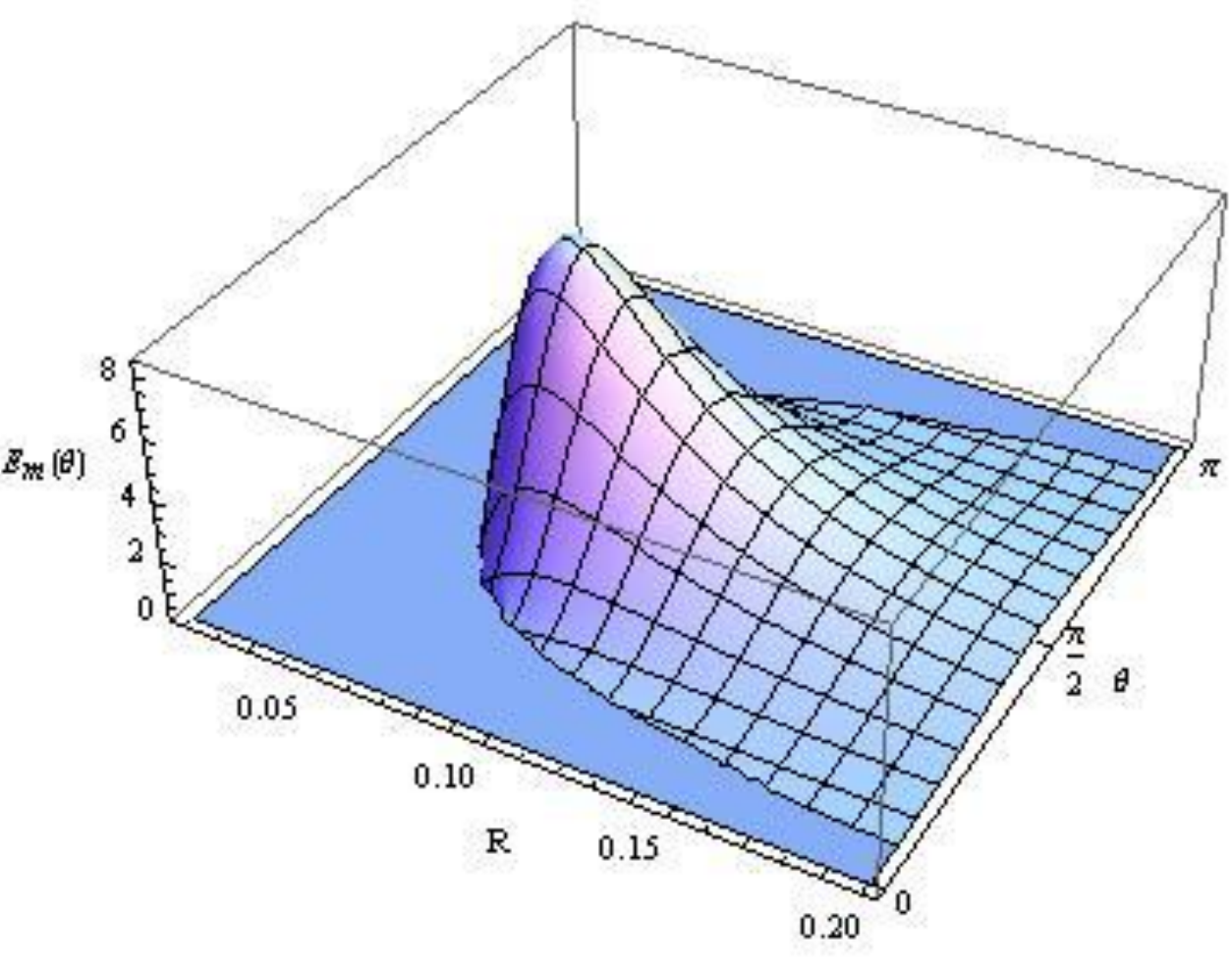}\\
\includegraphics[width=80mm]{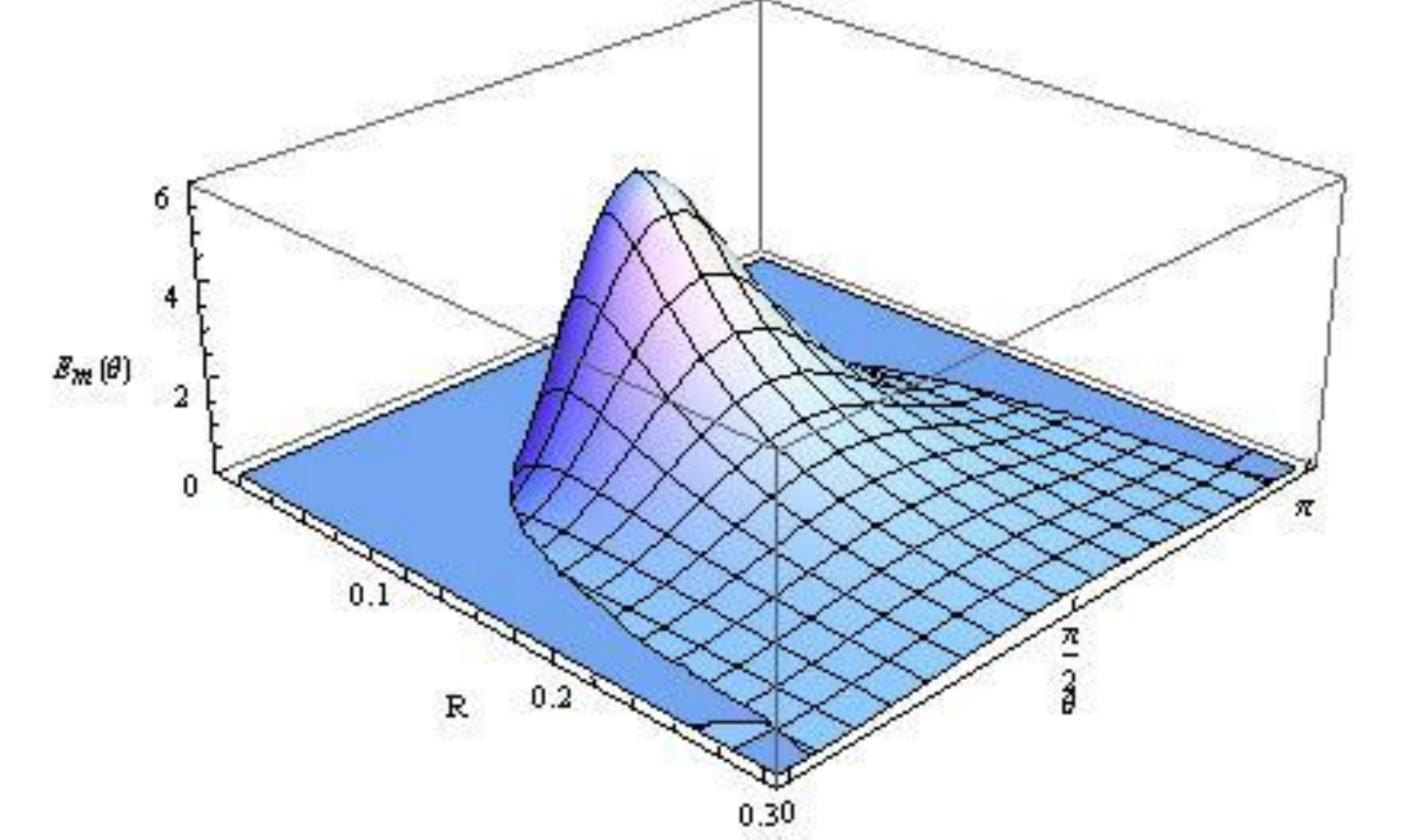}\\
\includegraphics[width=80mm]{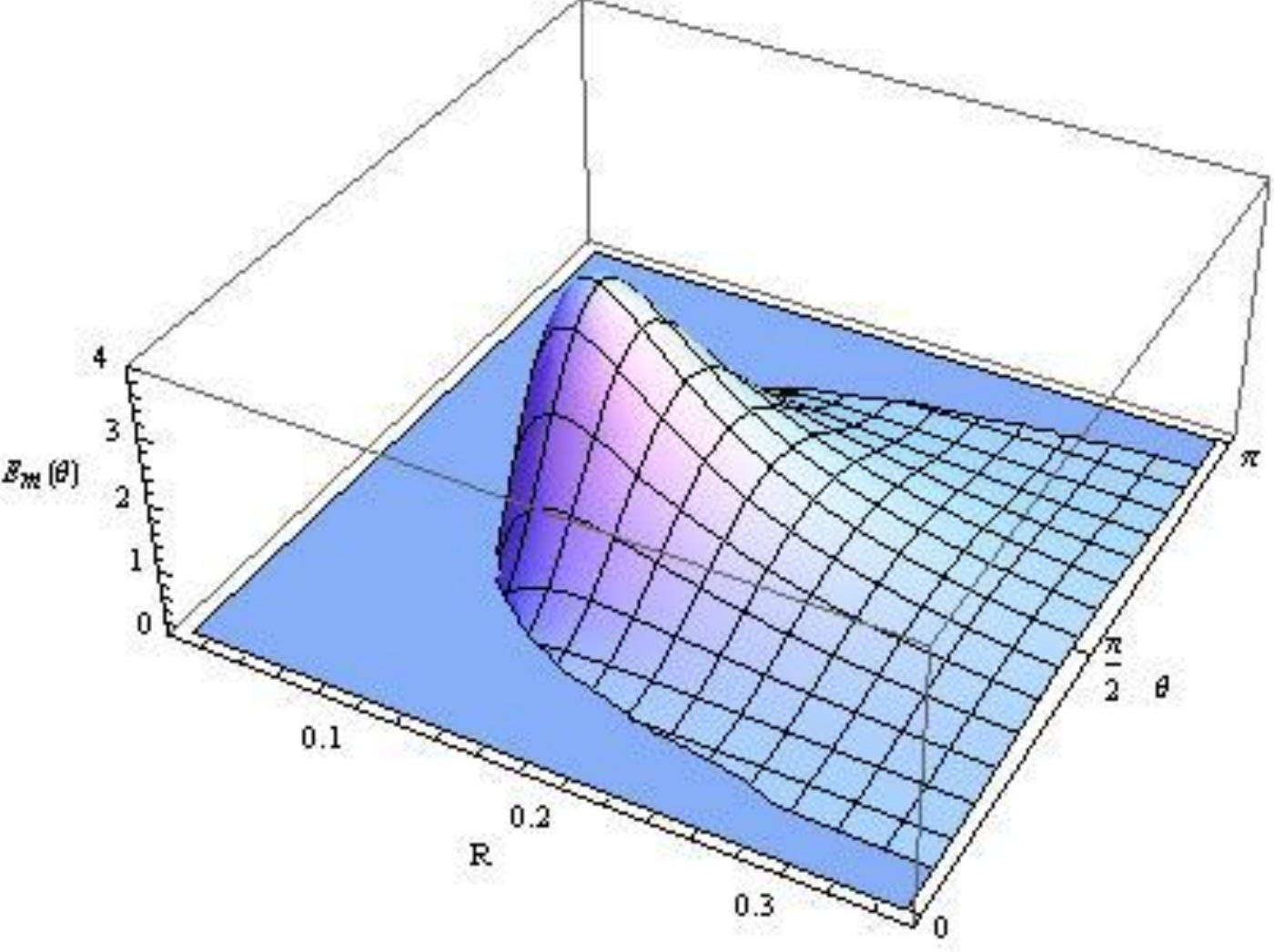}\\
\includegraphics[width=80mm]{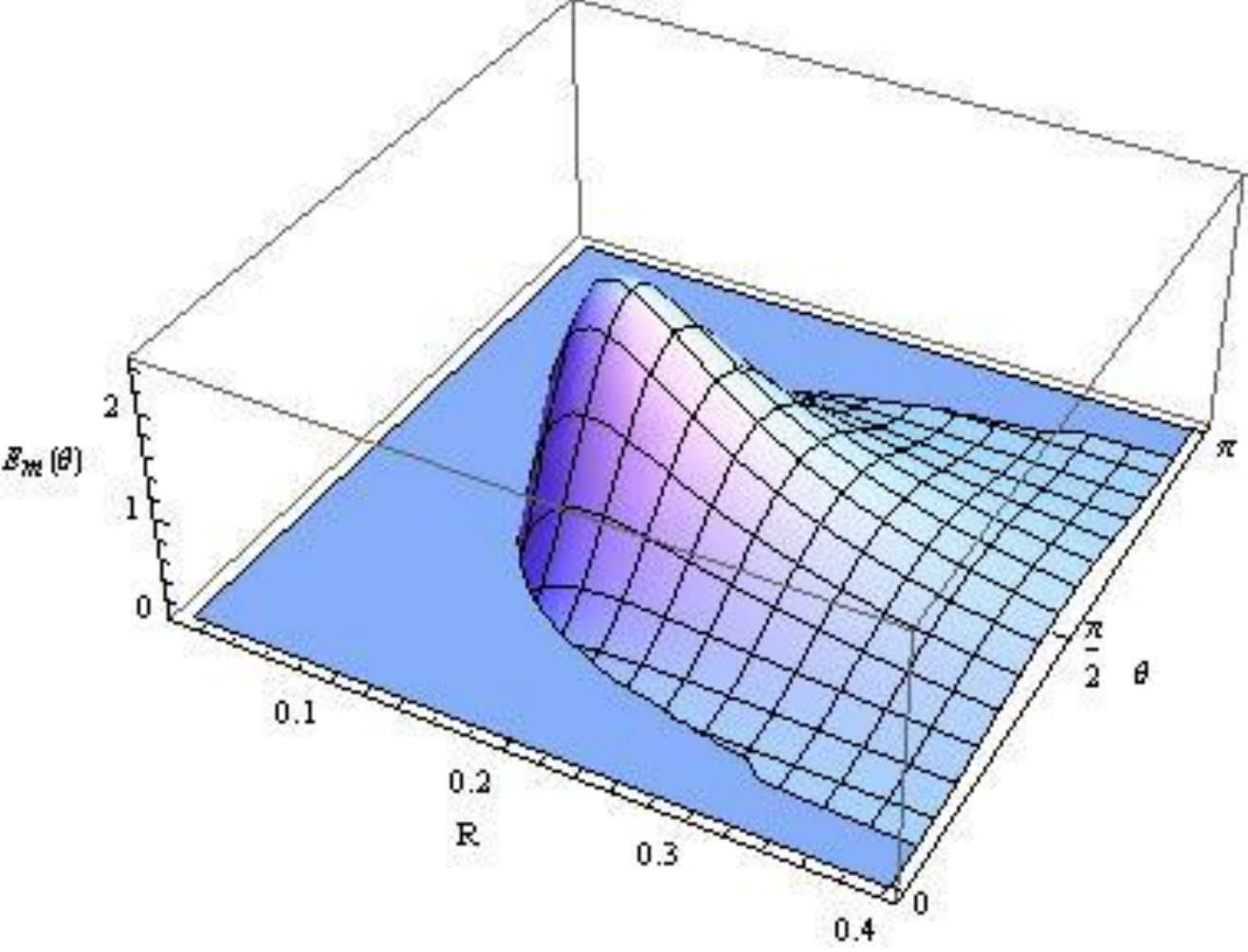}\\
\end{tabular}
\caption{(Color online) 3d-profiles of color electric field $E_{m}(\theta)$ for $\alpha_{s}=0.12,~0.24~0.48~and~0.96$ as a function of $R$ and $\theta$.}
\end{figure}  

\section{Thermal Effects on field configurations and associated critical parameters}
It has been argued that the multi flux tube structure of dual QCD leads to a viable explanation for the low-energy confining features of QCD and may further be used for exploring the phase structure of QCD under some unusual conditions like those of high temperatures and high densities. The behavior of QCD at finite temperature is, in fact, expected to play a vital role in understanding the dynamics of the QCD phase transition including QGP phase of nuclear matter \cite{hcc5}. Hence, in view of these facts, starting from the Lagrangian (5), let us use the partition functional approach alongwith the meanfield treatment for the QCD monopole field to evaluate the thermal contributions to the effective potential in the dual QCD. The partition functional, for the present dual QCD in the thermal equilibrium at a constant temperature $T$, may be given by an Eucledian path integral over a slab of infinite spatial extent and $\beta$ ($\equiv T^{-1}$) temporal extent, as,\\
\begin{equation}
Z[J]=\int D[\phi]D[B_{\mu}^{(d)}]exp(-S^{(d)}),
\end{equation}
where, $S^{(d)}$ is the dual QCD action and is given by,\\
\begin{equation}
S^{(d)}=-i\int d^{4}x(\mathcal{L}_{d}^{(m)}-J|\phi|^{2})
\end{equation}
where $\mathcal{L}^{(m)}_{d}$ is given by equation (5) and for the phase transition study in dual QCD vacuum the effective potential reliable in relatively weak coupling in near infrared regime is naturally desired as given in the form of equation (7). Under the thermal evolution of the QCD system, there are marked fluctuations in the monopole field and the effective potential at finite temperatures then corresponds to the thermodynamical potential which leads to the vital informations for the QCD phase transition. The thermal evolution of the QCD system as investigated \cite{hcc5} by taking into account the Dolan and Jackiw approach \cite{DJ} using high temperature expansion for the effective potential, ultimately leads to,\\
\begin{equation}
V_{eff}(\phi,T)=3\lambda\alpha_{s}^{-2}(\phi^{2}-\phi_{0}^{2})^{2} - \frac{7}{90}\pi^{2}T^{4}+\left(\frac{4\pi\alpha_{s} + \lambda}{2\alpha_{s}^{2}}\right)T^{2}\phi^{2}.
\end{equation}
It's minimization, in turn, leads to the thermally evolving vector glueball masses as,\\
\begin{equation}
m_{B}^{(T)} = (8\pi\alpha_{s}^{-1})^{1/2}\phi_{0}\biggl[1-\biggl(\frac{T}{T_{c}}\biggr)^{2}\biggr]^{1/2},
\end{equation}
where, the critical temperatue of phase transition as obtained by vanishing coefficient of terms quadratic in $\phi$ in effective potential (equation 41) is given as (for $\lambda =1$),\\
\begin{equation}
T_{c}=2\phi_{0}\sqrt{\frac{3}{4\pi\alpha_{s}+1}},
\end{equation}
which for different values of the couplings (i.e. for 0.12, 0.24, 0.48 and 0.96) leads to the critical temperatures as 0.318 GeV, 0.272 GeV, 0.220 GeV and 0.172 GeV respectively. With these considerations for the thermal evolution of the QCD system, the color electric field as given by equation (38) evolves at finite temperature in the following form,\\
\begin{equation}
E_{m}^{(T)}(\theta)=\tilde{E}_{m}^{(T)} exp(-Rm_{B}^{(T)} sin\theta)
\end{equation}
where, \begin{center}
$\tilde{E}_{m}^{(T)} = \frac{nC_{T}(4\pi\alpha_{s})^{1/2}}{8\pi R^{3/2} sin^{3/2}\theta}[1-2Rm_{B}^{(T)}sin\theta]$
\end{center}
with, $C_{T} = 2B(\sqrt{18}\lambda\pi)^{1/2}\alpha_{s}^{-5/2}(m_{B}^{(T)})^{1/2}$. The critical value of such color electric field around phase transition boundary may also be obtained using equation (36) in the following form,\\
\begin{equation}
E_{m}^{c}(T) = E_{m}^{c}[1-(T/T_{c})^{2}]
\end{equation}
In addition, the associated critical radius of phase transition and the critical flux tube density inside the hadronic sphere under high temperature in thermal QCD reduce to the following form,\\
\begin{equation}
R_{c}(T)=R_{c}\biggl(1-\frac{T^{2}}{T_{c}^{2}}\biggr)^{-1/2}~~and~~~d_{c}(T)=d_{c}\biggl(1-\frac{T^{2}}{T_{c}^{2}}\biggr),
\end{equation}
which show a large reduction in critical color electric field and the flux tube density in the phase transition region around critical temperature point and transition of the system to deconfined phase of large critical radii. The typical thermal profiles of color electric field as given by equation (44) have been depicted in figure 4, 5, 6 and 7 for different couplings in infrared sector of QCD. A considerable reduction in the field at increasing temperatures has been demonstrated by both 2-d and 3-d graphics. At temperatures, $T \geq T_{c}$, the field tends to get distributed uniformally around the centre of the sphere while it gets localized around the two poles ($\theta$~=~0~and~$\pi$) and drops down with its minimum around $\theta = \pi/2$. In any case, the system maintains perfect reflection symmetry around $\theta = \pi/2$ plane. For the physically accessible infrared sector ($\alpha_{s}=0.12$), the phase transition is expected around 0.235 GeV around which the flux tube density approaches to its minimum value and the magnetic condensate is evaporated into thermal monopoles.\\

\begin{figure}[htp]
\centering
\begin{tabular}{cc}
\includegraphics[width=80mm]{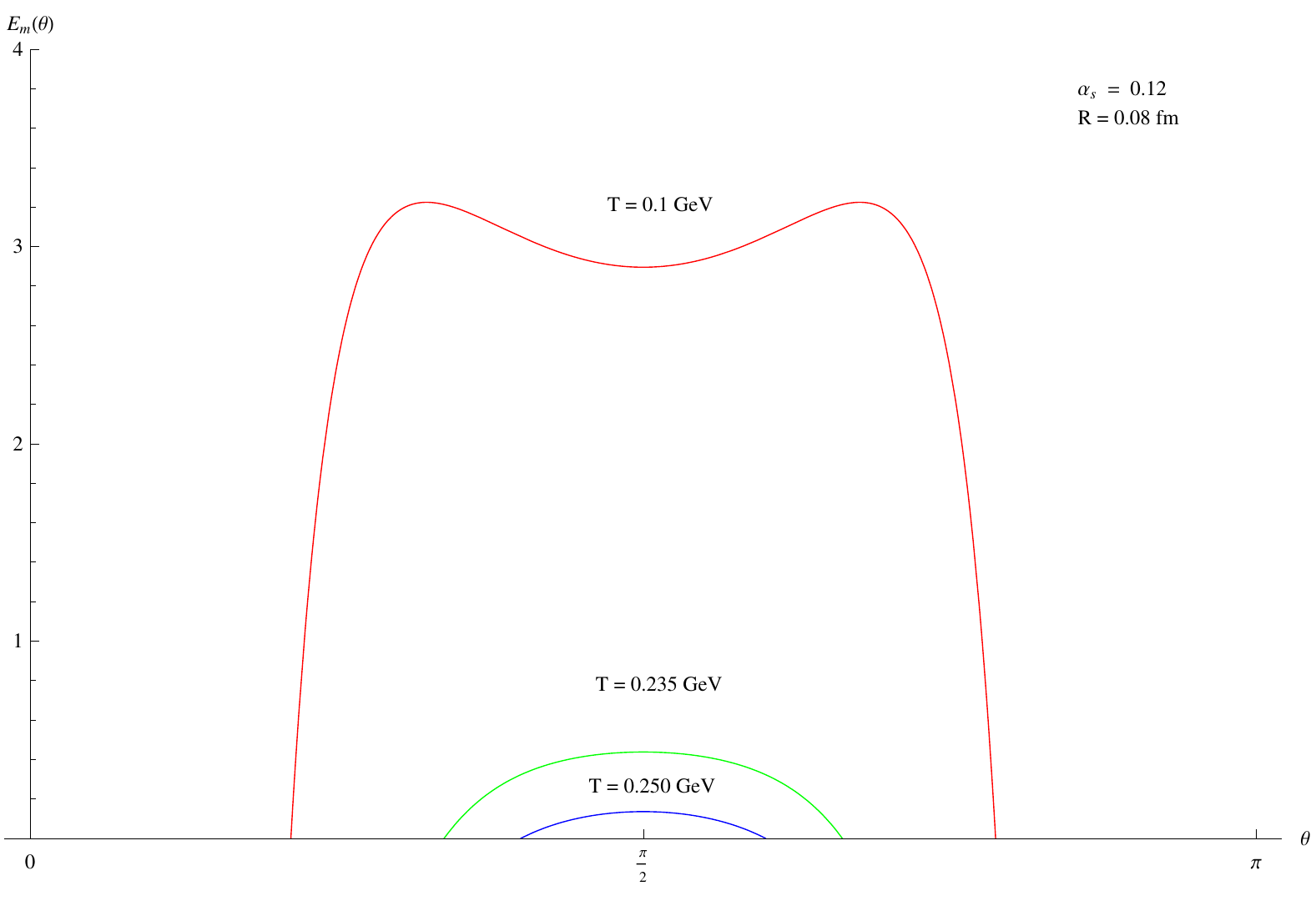}\\
\includegraphics[width=80mm]{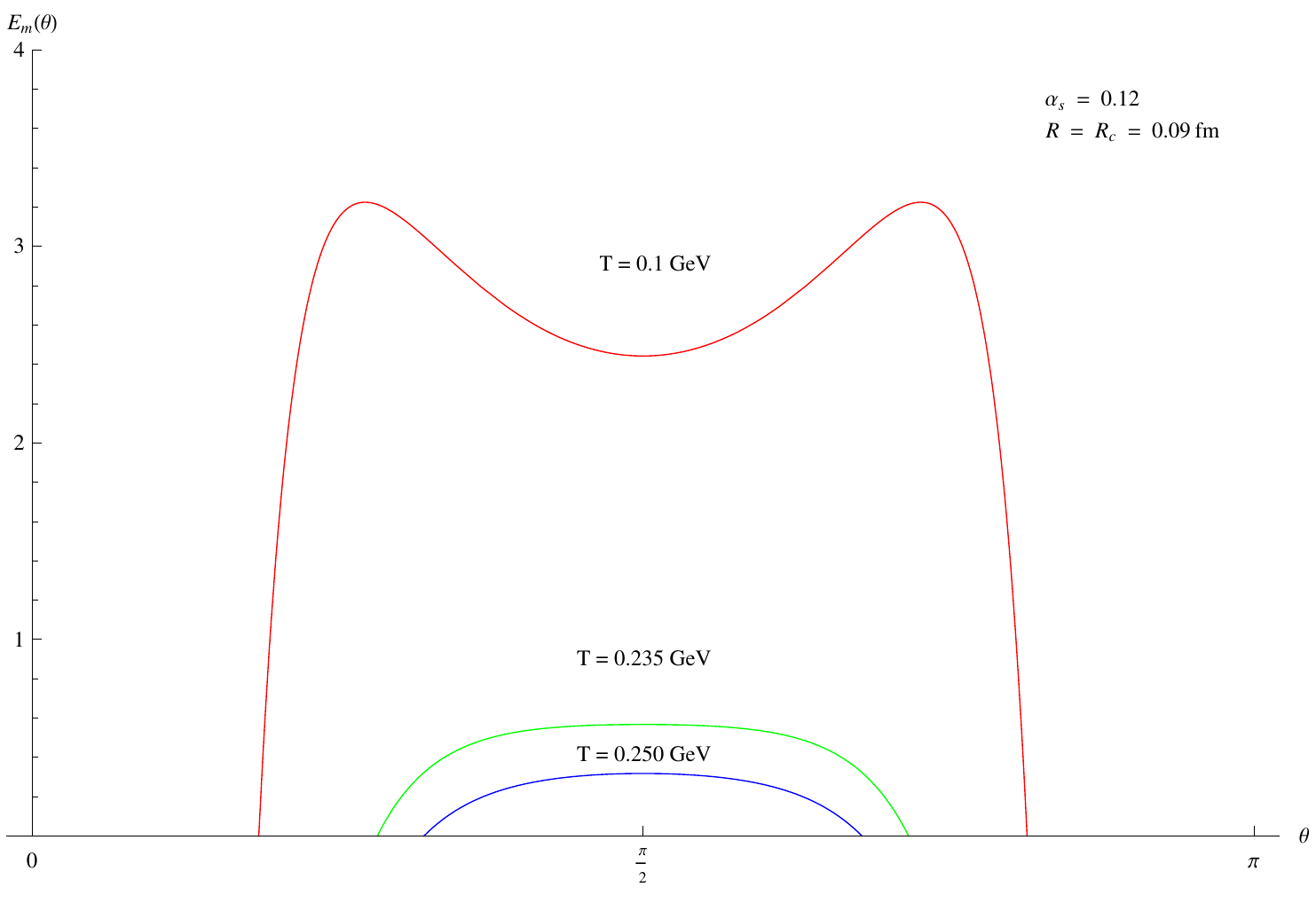}\\
\includegraphics[width=80mm]{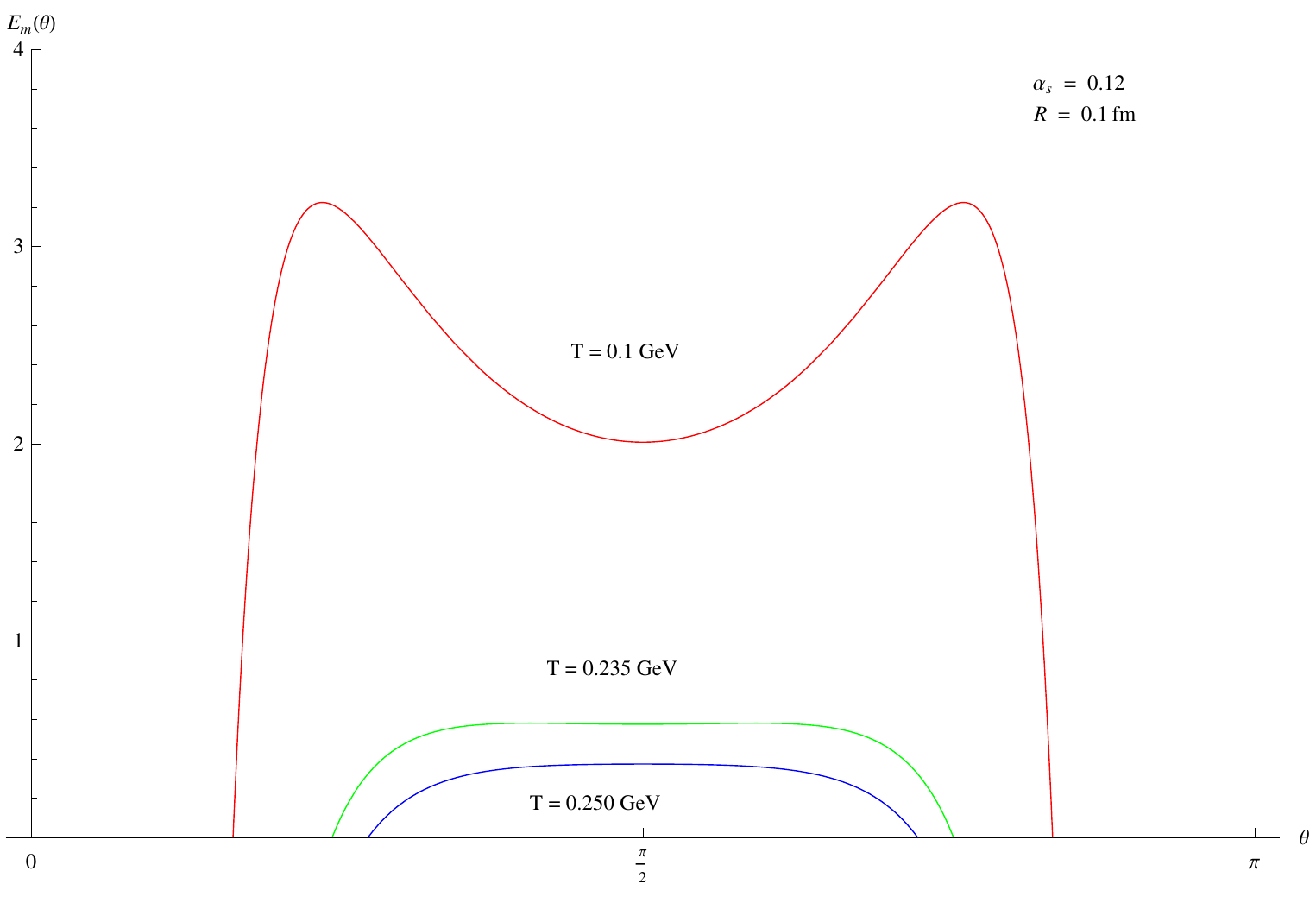}\\
\includegraphics[width=80mm]{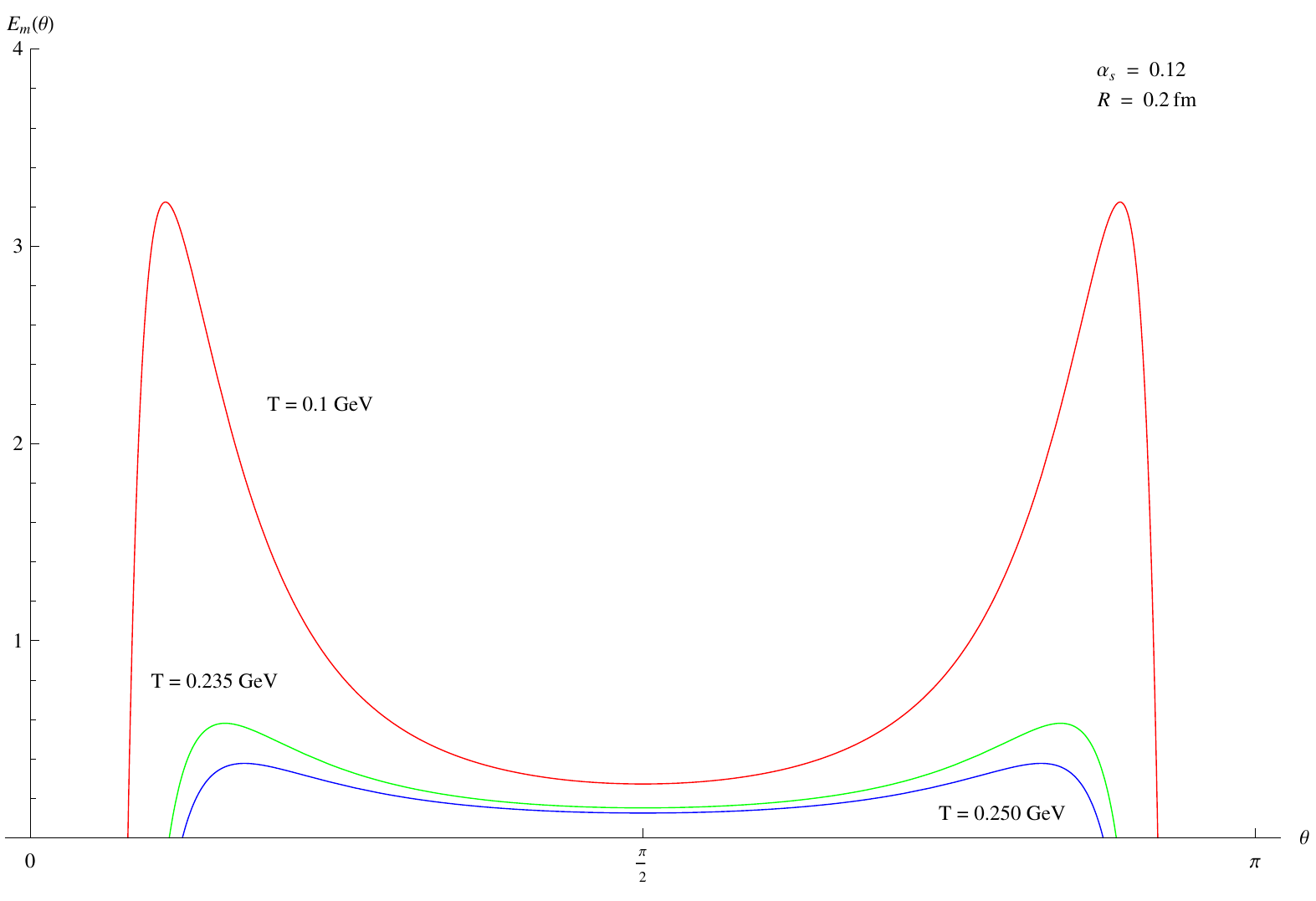}\\
\end{tabular}
\caption{(Color online) Profiles of color electric field $E_{m}(\theta)$ as a function of $\theta$ for different $T$ at $\alpha_{s}=0.12$}
\end{figure}

\begin{figure}[htp]
\centering
\begin{tabular}{cc}
\includegraphics[width=80mm]{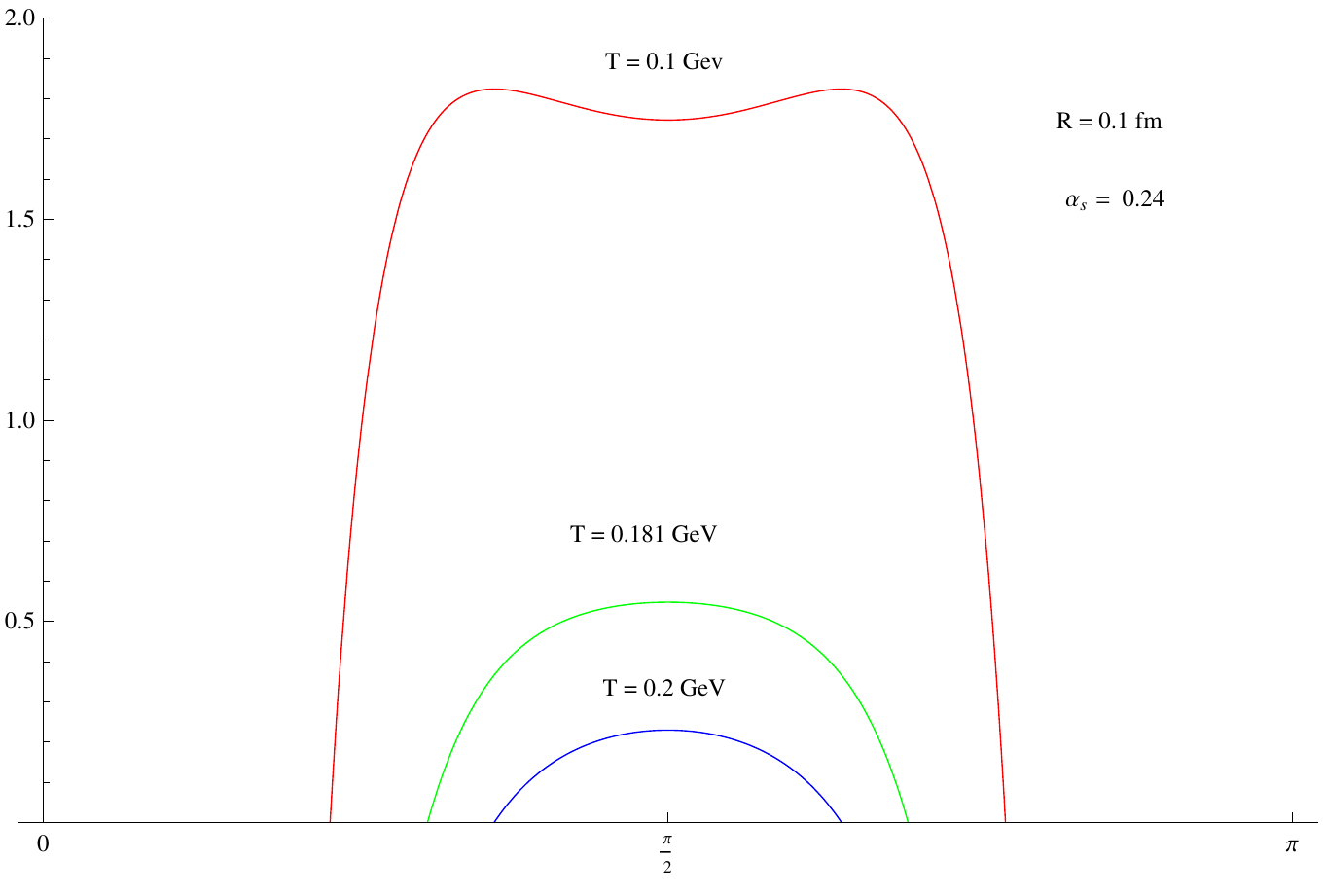}\\
\includegraphics[width=80mm]{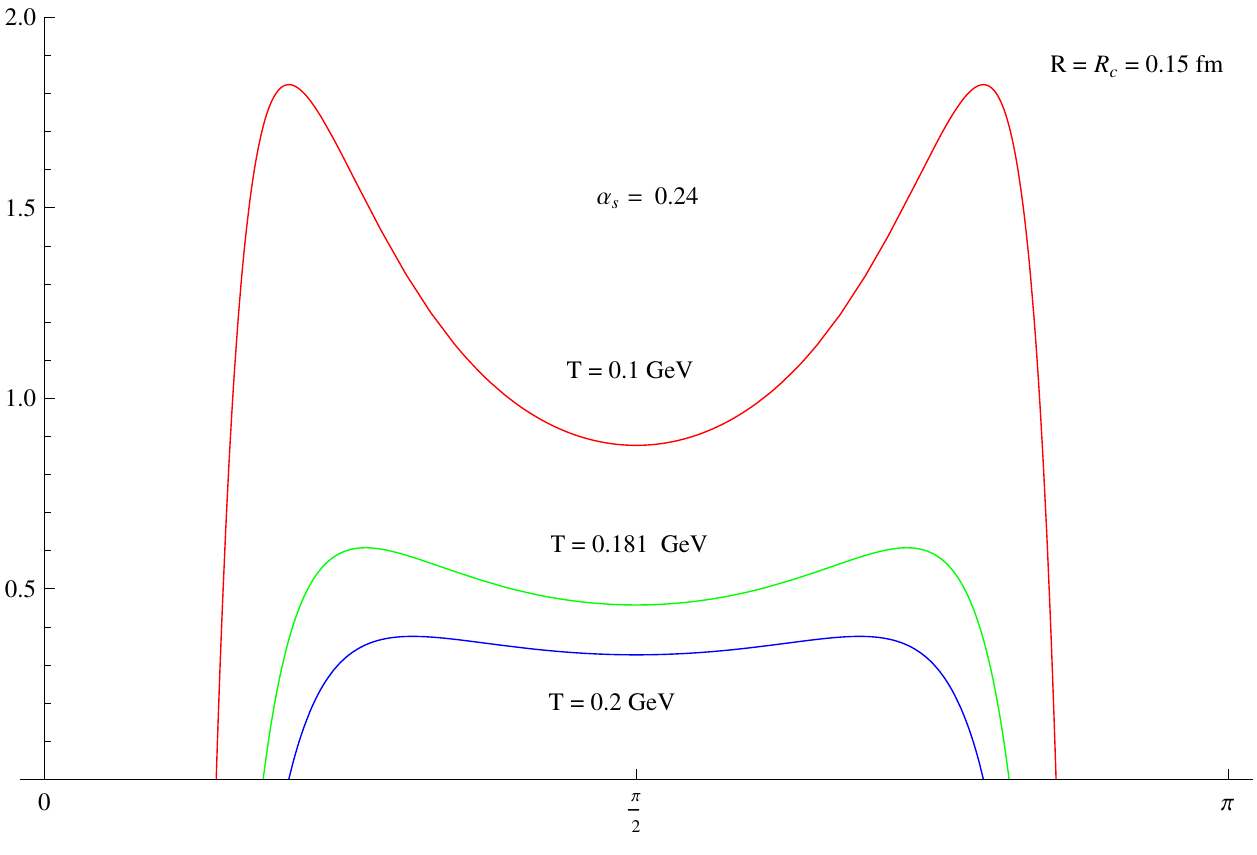}\\
\includegraphics[width=80mm]{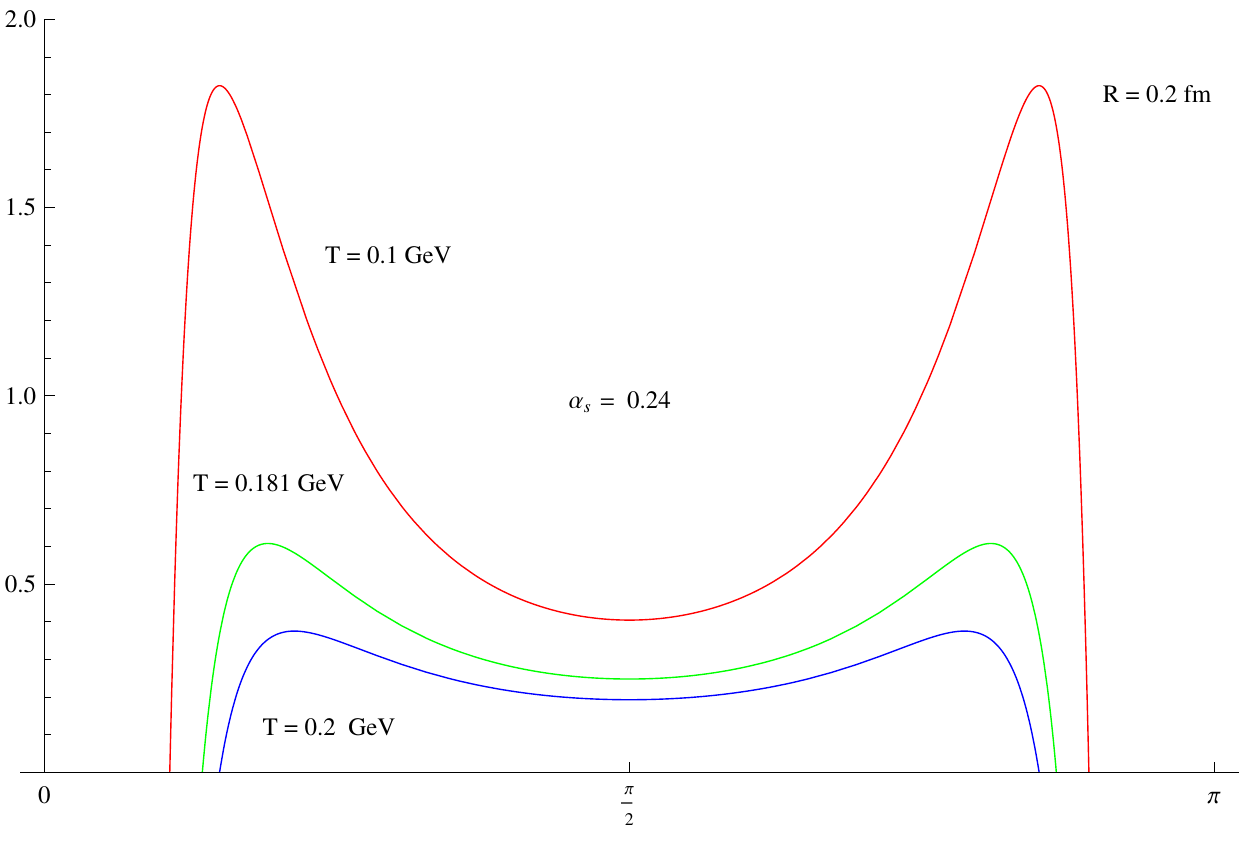}\\
\includegraphics[width=80mm]{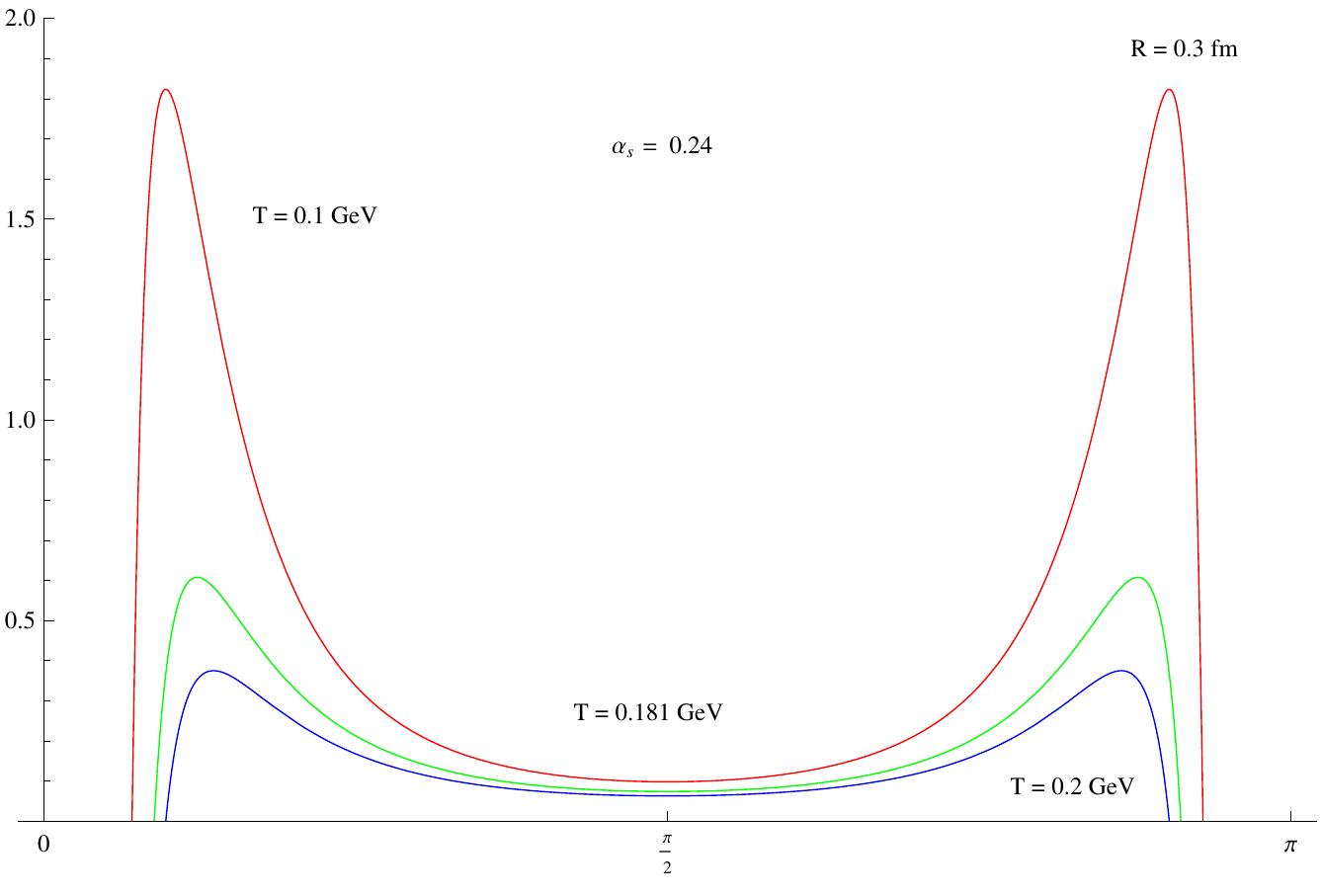}\\
\end{tabular}
\caption{(Color online) Profiles of color electric field $E_{m}(\theta)$ as a function of $\theta$ for different $T$ at $\alpha_{s}=0.24$}
\end{figure}

\begin{figure}[htp]
\centering
\begin{tabular}{cc}
\includegraphics[width=80mm]{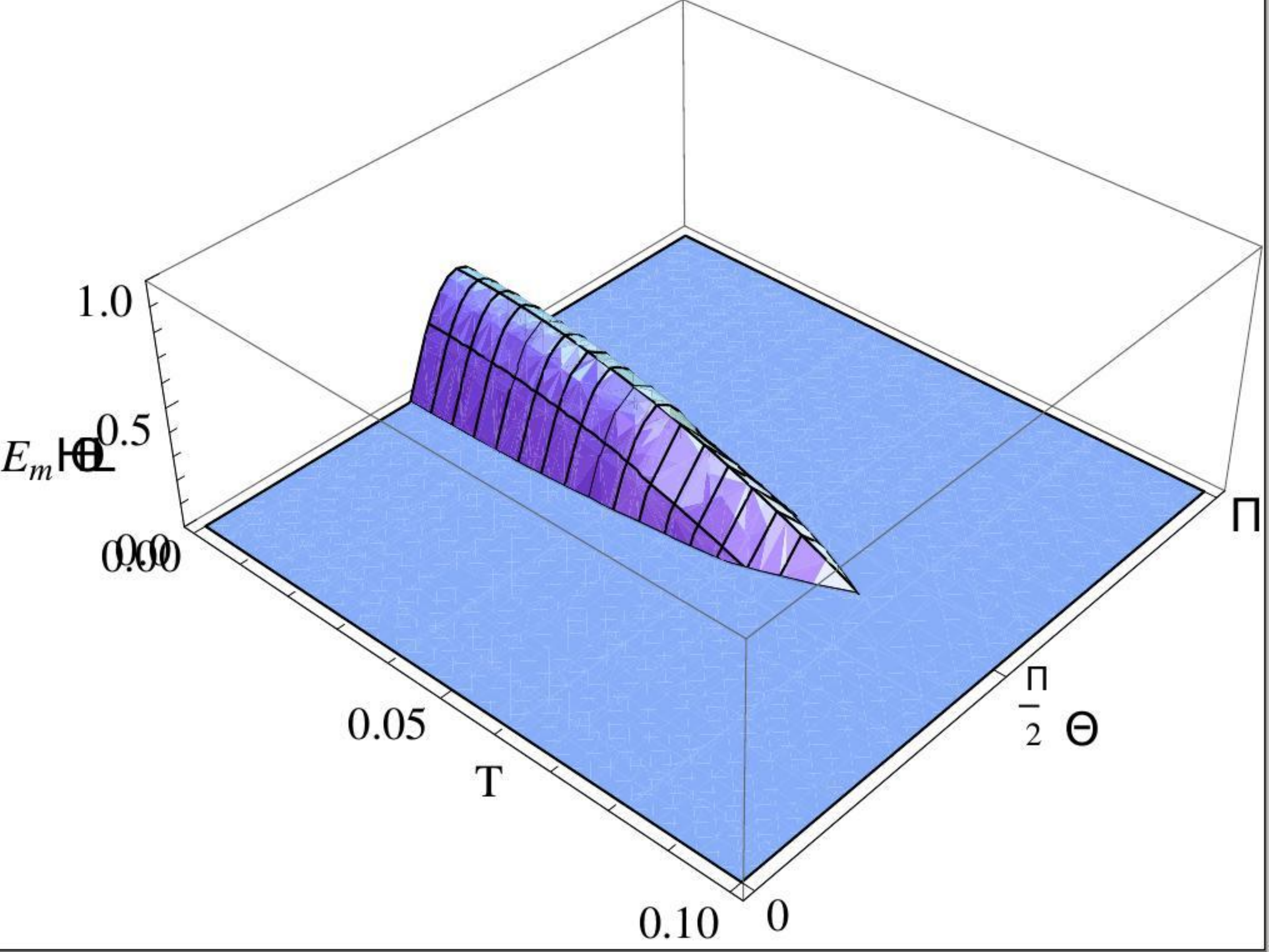}\\
\includegraphics[width=80mm]{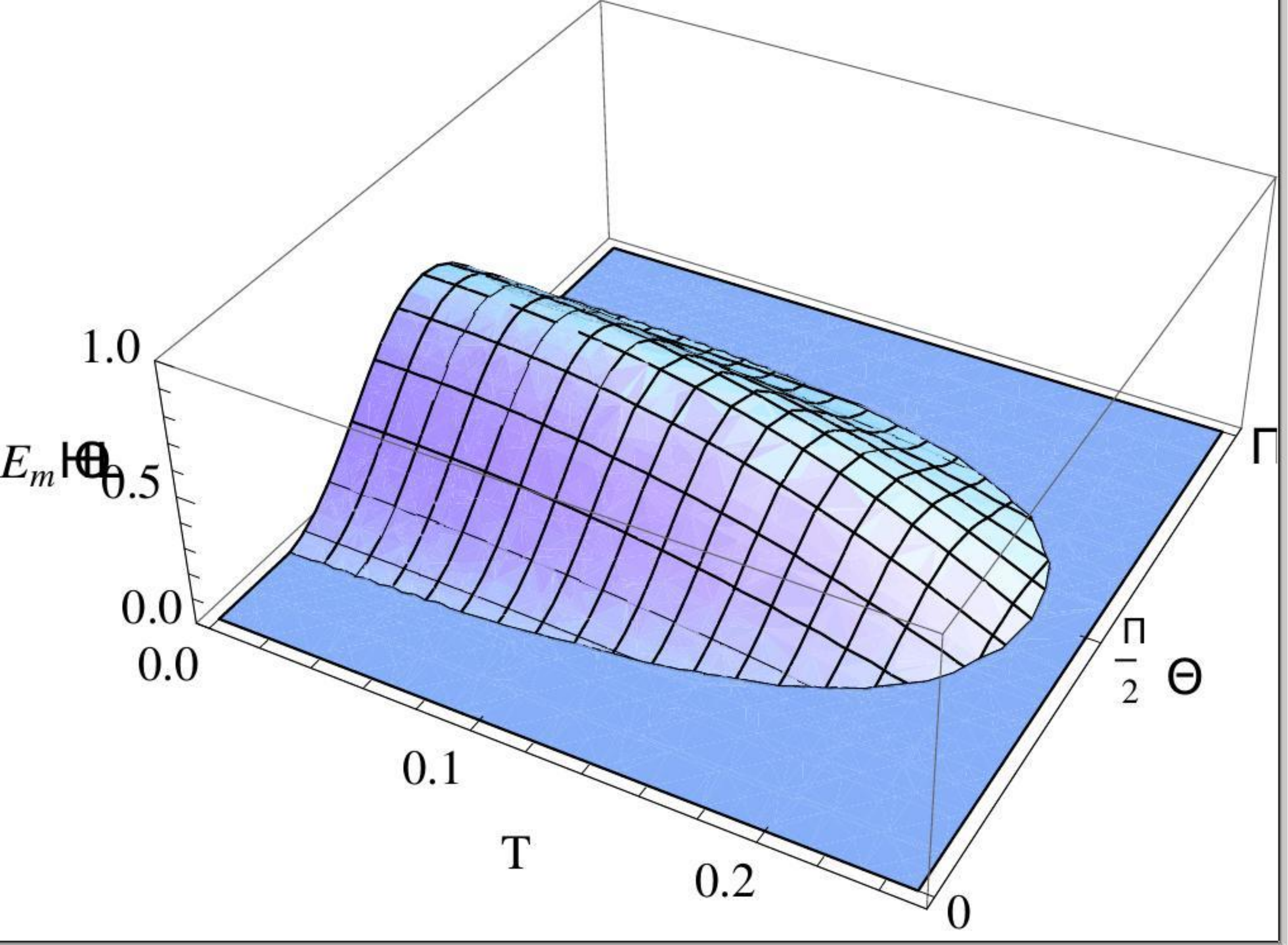}\\
\includegraphics[width=80mm]{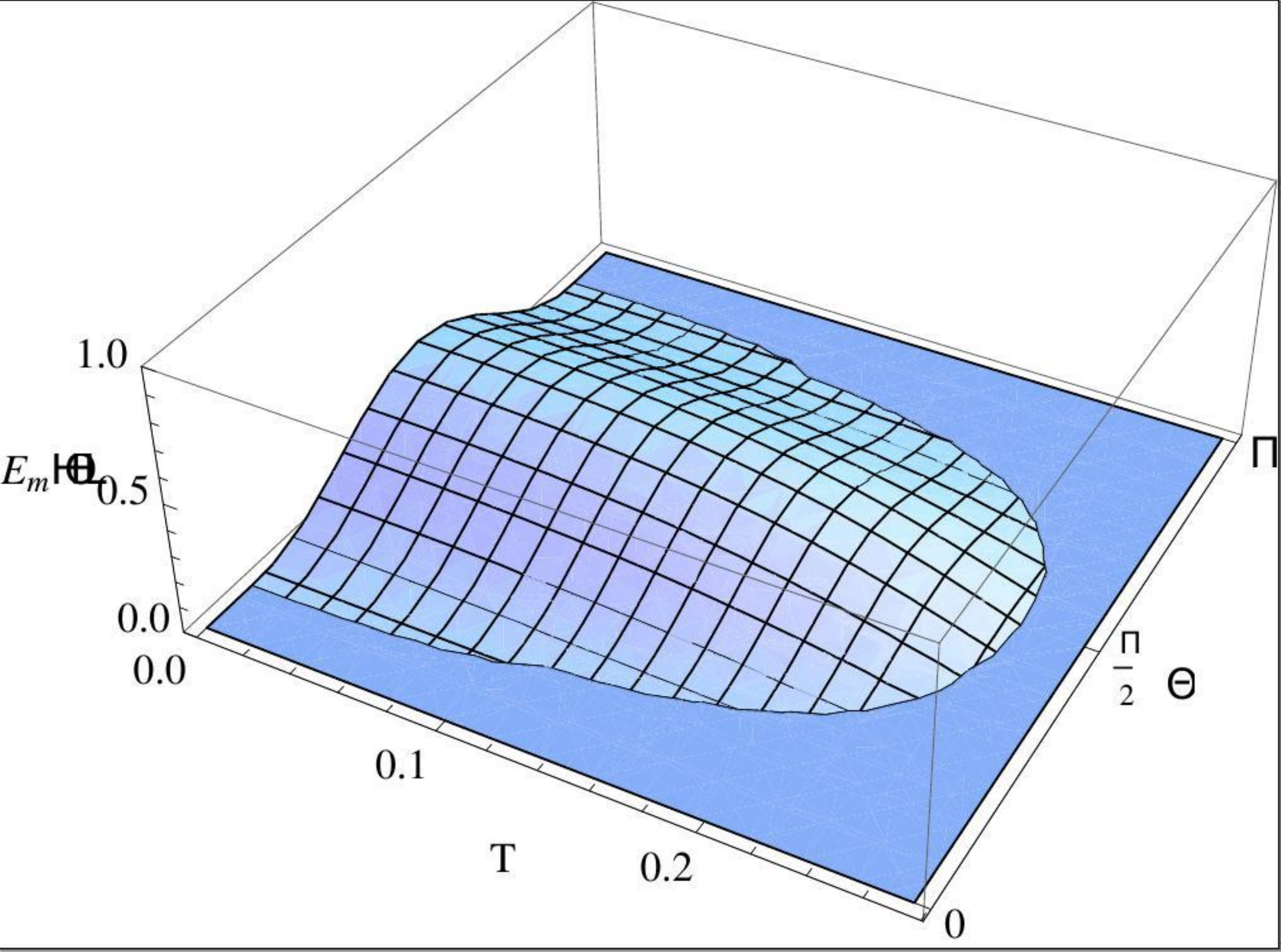}\\
\end{tabular}
\caption{(Color online) Thermal variation of the profiles of color electric field $E_{m}(\theta)$ for $\alpha_{s}=0.12$ at $R$ = 0.1, 0.2 and 0.25 fm respectively.}
\end{figure}

\begin{figure}[htp]
\centering
\begin{tabular}{cc}
\includegraphics[width=80mm]{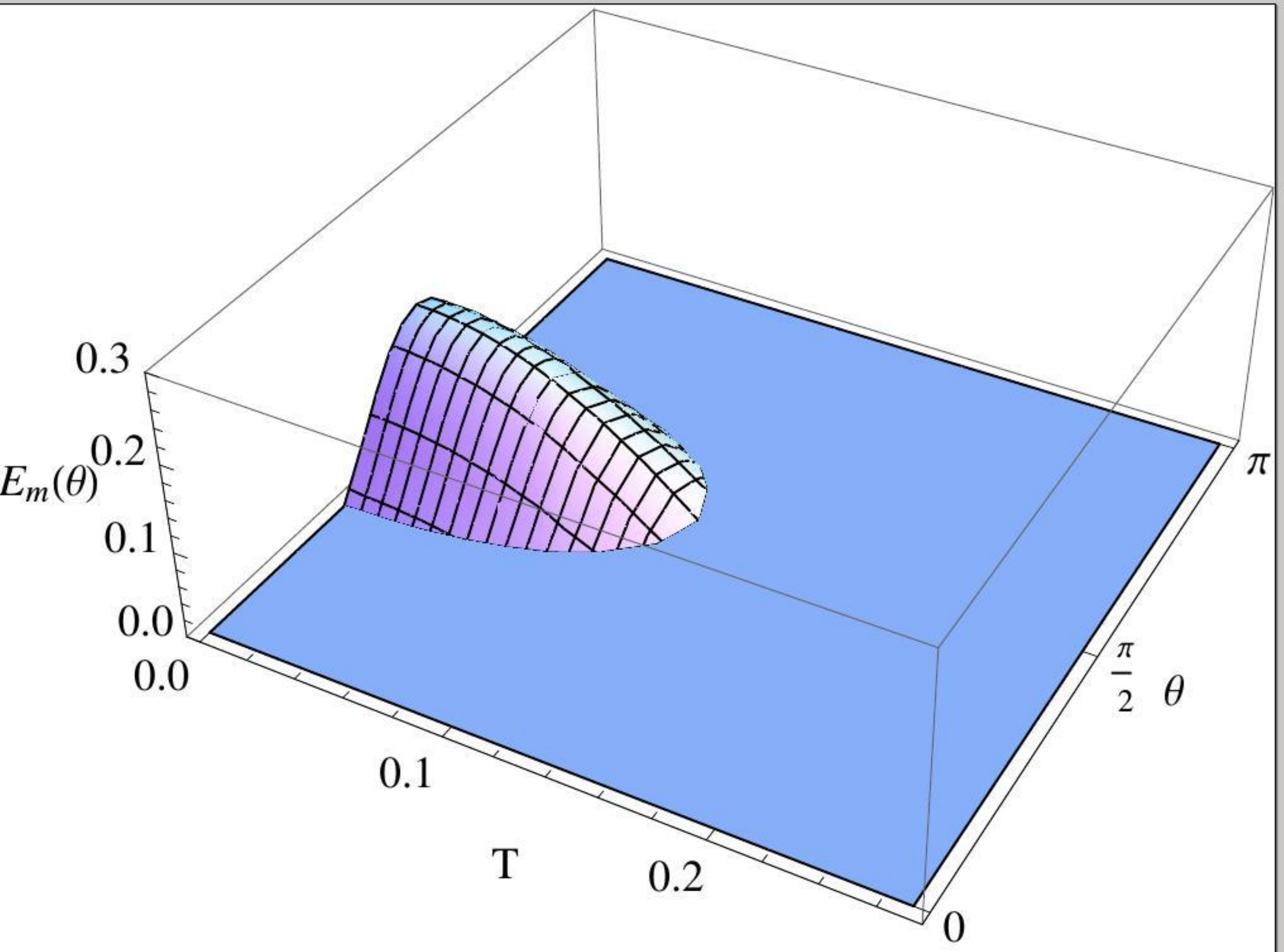}\\
\includegraphics[width=80mm]{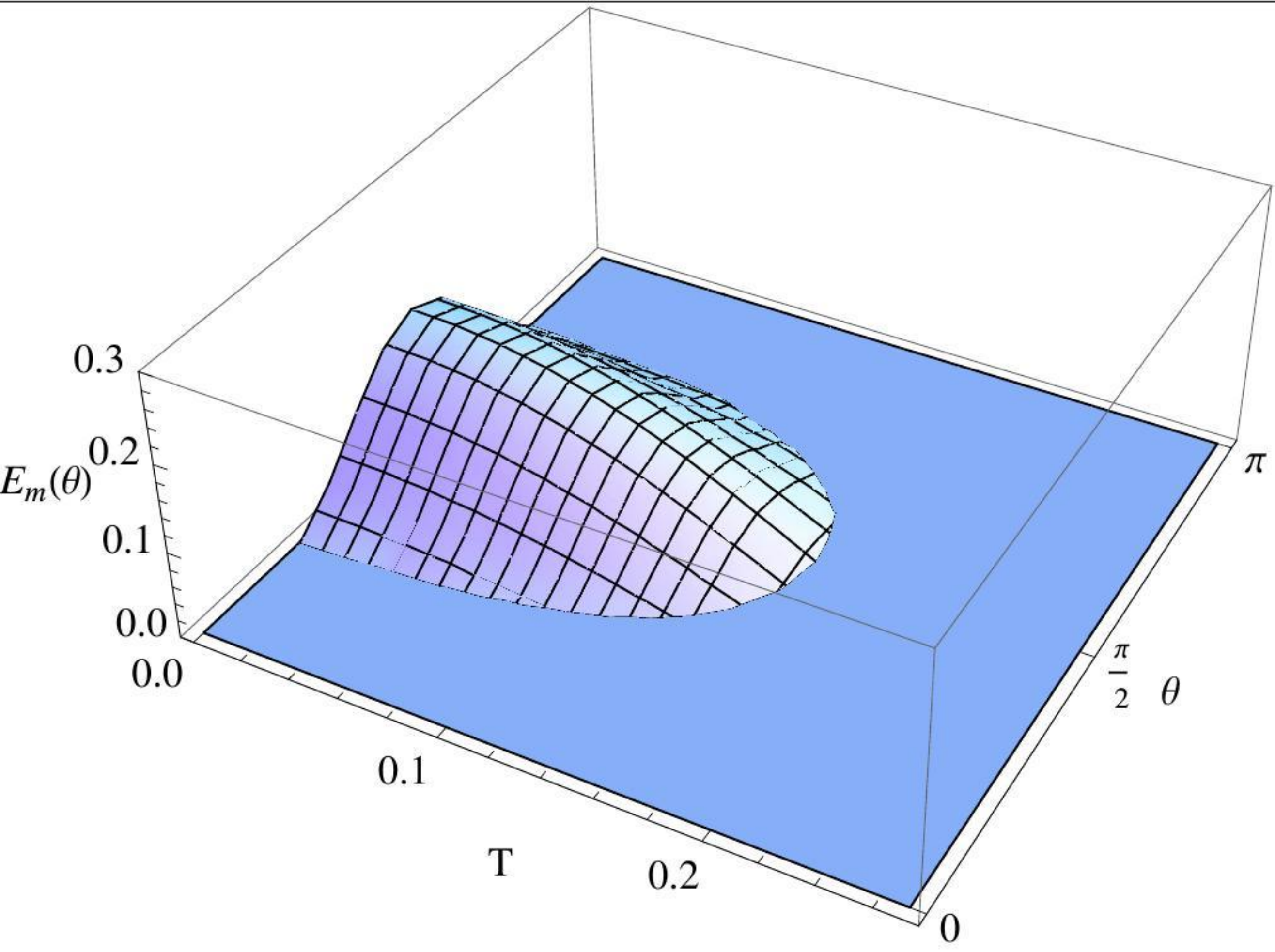}\\
\includegraphics[width=80mm]{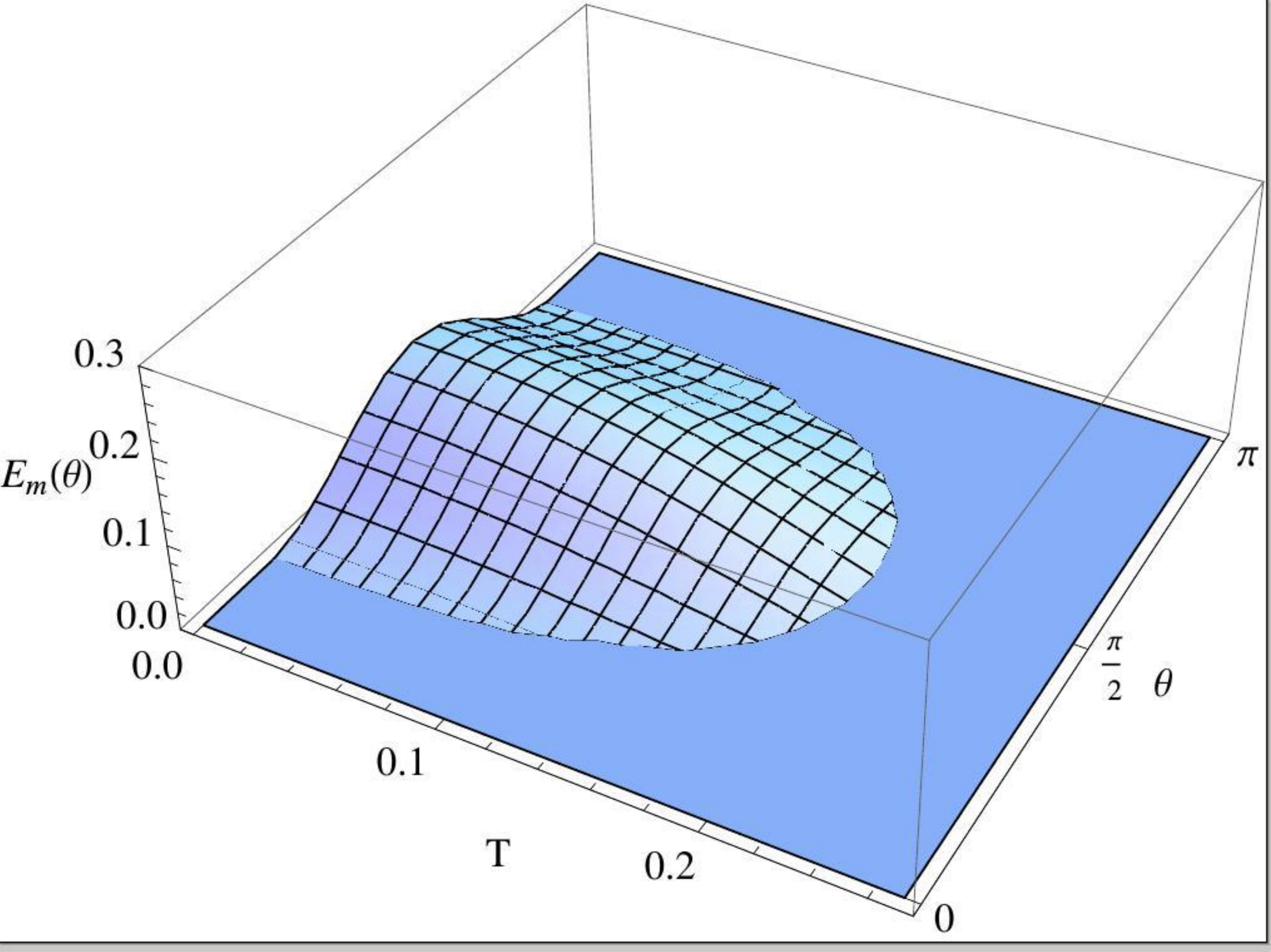}\\
\end{tabular}
\caption{(Color online) Thermal variation of the profiles of color electric field $E_{m}(\theta)$ at $\alpha_{s}=0.24$ for for different $R$ = 0.15, 0.2 and 0.25 fm respectively.}
\end{figure}

\section{Summary and Conclusions}
\hspace{0.5cm}In view of the fact that the topological degrees of freedom of non-Abelian gauge theories play a crucial role in understanding the various nonperturbative features as well as the underlying phase structure of QCD vacuum, the magnetic symmetry based dual version of QCD has been utilized which is shown to lead to a dual dynamics between color isocharges and topological charges imparting the dual superconducting properties to QCD on dynamical breaking of magnetic symmetry and confining any color isocharge  present. A unique periodically distributed flux tube structure has been shown to emerge in magnetically condensed QCD vacuum which has been analysed on a $S^{2}$ sphere to compute the critical parameters of phase transition using energy-balance condition for the energy components given by equations (19-27). The higher value of critical flux tube density around phase transition region in near infrared sector is shown to drive the system through QGP phase also. Further, the color electric flux quantization and the energy minimization conditions have been shown to lead to a uniformally distributed color electric field given by equation (35) with its critical value as given by equation (36) in the deconfinment region which immediately drops to more than ninety percent in the far-infrared sector of QCD. Using the asymptotic solutions for the dual potential, the analytic expression of the color-electric field has been derived as equation (38) and the profiles of color electric field at different hadronic scales and different values of coupling constant are depicted as 2-d and 3-d graphics of figures 2 and 3 respectively. It clearly demonstrate the localization of the color electric field around poles at large distance scales which goes homogeneous below the corresponding critical radius and confined in the central region that leads to the possibility of a homogeneous QGP formation due to increased flux-tube density before transiting to completely deconfined phase in the color screening region. The formation of this new kind of phase is expected as a result of the flux tube annihilation because of the increased flux tube density in the central region and, therefore, the flux tube density in dual QCD naturally plays a key role in creation of QGP in its infrared sector. It is important to note that, in the present scenario, below the critical radius the flux tube density does not completely vanish in the color screening regime which, in turn, strongly support the formation of an intermediate phase before transiting to the completely deconfined region. In the near infrared region physically relevant for the phase transition, the critical flux-tube density has been shown to be in agreement with that computed for Pb-Pb central heavy-ion collisions and demonstrates the feasibility of multi-flux tube formulation of dual QCD in the study of phase structure of QCD. Further, the distribution of color electric field at or few separation below the critical radius still indicates the survival of color flux tubes which points out the stability of QCD vacuum as well as the strongly interacting \cite{bk}-\cite{heinz} behavior of such intermediatery phase in QCD.\\
Furthermore, since the thermal effects are extremely important and play a dominant role in phase transition as well as QGP formation process, the thermal evolution of color electric field has also been investigated for different hadronic scales which leads to the weakening of confining force with temperature and is expected to trigger the flux tube melting before transiting to pure asymptotically free phase of hadronic matter. In the thermal environment, as the temperature is increased and approached to its critical value, the amplitude of the field in the interior of flux tube gets supressed but still supports the survival of flux tubes upto some extent beyond the critical temperature in QCD vacuum which characterizes the intermediatory and weakly interacting clusters of colored particles. On relatively higher temperature scales, it resembles with the evaporation of flux tubes and subsequently the transition of system into completely deconfined state. Such thermalization process with quarks and gluons and possibility of QGP formation then become quite similar to that of the heavy-ion collisions where several interacting flux tubes are expected to overlap in the central region and lead to the similar kind of changes in the shape of the flux tube as a result of their annihilation or unification and pushes the system into either QGP phase or passes through the crossover.\\ 
  
\section{Acknowledgments}

One of the authors, D.S. Rawat, is thankful to the University Grants Commission, New Delhi, for financial assistance in the form of a RFSMS fellowship. The authors (DSR) is also thankful to Prof. B.K. Patra, I.I.T. Roorkee, India to invite DSR at I.I.T. Roorkee under TPSC programme for useful discussion.

\end{document}